\documentclass{jpp}
\usepackage{graphicx}
\usepackage{natbib}

\ifCUPmtlplainloaded \else
  \checkfont{eurm10}
  \iffontfound
    \IfFileExists{upmath.sty}
      {\typeout{^^JFound AMS Euler Roman fonts on the system,
                   using the 'upmath' package.^^J}%
       \usepackage{upmath}}
      {\typeout{^^JFound AMS Euler Roman fonts on the system, but you
                   dont seem to have the}%
       \typeout{'upmath' package installed. JPP.cls can take advantage
                 of these fonts, if you use 'upmath' package.^^J}%
      }
  \else
  \fi
\fi

\ifCUPmtlplainloaded \else
  \checkfont{msam10}
  \iffontfound
    \IfFileExists{amssymb.sty}
      {\typeout{^^JFound AMS Symbol fonts on the system, using the
                'amssymb' package.^^J}%
       \usepackage{amssymb}%
       \let\le=\leqslant  
       \let\ge=\geqslant  
      }{}
  \fi
\fi

\ifCUPmtlplainloaded \else
  \IfFileExists{amsbsy.sty}
    {\typeout{^^JFound the 'amsbsy' package on the system, using it.^^J}%
     \usepackage{amsbsy}}
    {\providecommand\boldsymbol[1]{\mbox{\boldmath $##1$}}}
\fi

\newsavebox{\astrutbox}
\sbox{\astrutbox}{\rule[-5pt]{0pt}{20pt}}

\def\ni{\noindent}

\def\emfb{\overline{\mbox{\boldmath ${\cal E}$}} {}}

\def\beq{\begin{equation}}
\def\ee{\end{equation}}
\def\eeq{\end{equation}}
\def\lsim{\mathrel{\rlap{\lower4pt\hbox{\hskip1pt$\sim$}}
    \raise1pt\hbox{$<$}}}
\def\gsim{\mathrel{\rlap{\lower4pt\hbox{\hskip1pt$\sim$}}
    \raise1pt\hbox{$>$}}}

\def\bfB{{\bf B}}

\def\lb{\langle}
\def\rb{\rangle}

\def\bfv{{\bf v}}

\def\bfb{{\bf b}}

\def\bfB{{\bf B}}

\def\bbB{\overline {\bf B}}

\def\div{\nabla\cdot}

\title[Challenges and directions for next generation  accretion disc theory]{
 Motivation and challenge  to capture both  large scale  and local transport in next generation accretion theory}

\author[E. G. Blackman and F. Nauman]%
{E\ls R\ls I\ls C\ls \ G.\ \ls B\ls L\ls A\ls C\ls K\ls M\ls A\ls N$^{1,2}$%
  \thanks{IBM-Einstein Fellow; Simons Fellow; Email address for correspondence: blackman@ias.edu}\ns
F\ls A\ls R\ls R\ls U\ls K\ls H\ls \ N\ls A\ls U\ls M\ls A\ls N$^1$}

\affiliation{$^1$Department of Physics and Astronomy, University of Rochester,
Rochester, NY, 14627, USA\\[\affilskip]
$^2$School of Natural Sciences, Institute for Advanced Study,
Einstein Drive, Princeton NJ, 08540, USA}

\pubyear{2010}
\volume{650}
\pagerange{119--126}
\date{?; revised ?; accepted ?. - To be entered by editorial office}

\begin{document}

\maketitle

\begin{abstract}
Accretion disc theory is  less developed than  stellar evolution theory although  a similarly mature phenomenological picture is ultimately desired.  While  the interplay of theory and numerical simulations  has amplified community awareness of the role of magnetic fields in angular momentum transport,  there remains a long term challenge to incorporate  insight gained from  simulations back into improving  practical  models for comparison with  observations.  What has been learned  from simulations  that can lead to improvements beyond SS73 in practical models?  Here  we emphasize the need to incorporate the role of non-local transport more precisely.  
To show where large scale transport would fit into the theoretical framework and how it is currently missing,  we  review why the    wonderfully practical approach  of Shakura-Sunyaev (1973,SS73) is necessarily a mean field theory, and one which does not include  large scale transport.   Observations of coronae and jets combined with  the interpretation of results  even from shearing box simulations of the magnetorotational instability (MRI)  suggest that  a significant fraction of disc transport is indeed non-local.  We show that  the Maxwell stresses in saturation are dominated by large scale contributions and the physics of MRI transport is not  fully captured by a  viscosity. 
We also clarify the standard physical interpretation of the MRI as it applies to  shearing boxes.  
Computational limitations have so far focused  most attention toward local simulations  but  the next generation of global simulations should help to inform improved mean field  theories.   Mean field accretion theory and mean field dynamo theory should in fact be 
unified into  a single  theory that predicts  the time evolution of spectra and luminosity from separate disc, corona, and outflow contributions.   Finally, we note that any  mean field  theory, including that of SS73,  has a finite predictive precision that needs to be quantified when comparing the predictions to observations.

\end{abstract}

\begin{PACS}
Authors should not enter PACS codes directly on the manuscript, as these must be chosen during the online submission process and will then be added during the typesetting process (see http://www.aip.org/pacs/ for the full list of PACS codes)
\end{PACS}

\section{Introduction}
Building on earlier notions of \cite{Swedenborg1734},   \cite{Kant1755} (and independently  \cite{Laplace1796})
 proposed that the the sun and stars must have formed from material  initially rotating and subject to the influence of gravity. Early
scenarios could not explain the distribution of angular momentum in the solar system but collectively represented a 
step toward  accretion disc theory.   Two additional paths  toward modern accretion theory emerged in the 20th century from (1) the context of stars accreting from an ambient medium \citep{Hoyle1939,Bondi1952} and (2) from a  need to explain the extraordinary compact powers of quasars and extragalactic radio sources.  The latter observations stimulated explication of how the release  of copious  positive
kinetic or radiative energy can occur as  dense material falls deeper into a negative energy gravitational potential well   \citep{Shklovskii1963,Salpeter1964,Zeldovich1964}.  

The importance of  angular momentum transport (and thus discs) in this process, and the development of  specific hydrodynamic models  for close binary systems  \citep{Prendergast1968} along with magnetohydrodynamic (MHD) scenarios for  quasars  \citep{Lynden-Bell1969} soon emerged.  Shortly thereafter,  proteges of Zeldovich developed the now famous 1-D axisymmetric   $\alpha$ disc   formalism \citep{Shakura1973} connecting accretion  dynamics to   observed spectra.  This formalism has provided the 
most commonly  used practical framework for comparing  predicted accretion disc continiuum spectra to observations for accretion engines of all scales.
 
Essential for understanding  accretion discs is determining how  material can lose enough angular momentum
to fall deeper into the potential well and sustain the output power. 
This need  for angular momentum transport arises because  material that becomes an accretion engine typically originates from radii many orders of magnitude away.  If angular momentum is conserved upon infall,  the ratio of orbital speed to  Keplerian speed varies as the square root of the  distance to the central engine.  Therefore, 
material can fall in   conserving angular momentum only down to the radius at which the local Keplerian speed is reached. Inside of  this  radius, further infall requires a loss of angular momentum.   
 
The time scale for angular momentum loss determines  the accretion time scale.  When observed time scales from variability and/or system lifetimes are combined  with independent estimates of  disc surface densities, estimates of the efficacy of angular momentum transport can be made.  Dwarf novae (DN)  provide the most demanding constraints \citep{King2007,Kotko2012}. 
Commensurate with constraints from active galactic nuclei (AGN) and young stellar objects (YSOs), they demand  that the
angular momentum transport rate exceed, by many orders of magnitude,  that which microphysical  transport can provide.  
In recent decades,  an amusing  irony  has also emerged however:   all galaxies likely harbor central black holes but  only a small fraction have active engines.  Most are under-active, being either radiatively inefficient accreters or are being fueled with  very low accretion rates.
(See \cite{Yuan2014} for a review of radiatively inefficient accretion modes.)
The need for  mechanisms that lead to enhanced angular momentum transport in luminous sources
 is thus complemented by the need for  unusual quiescence in others, or  temporally evolving states within the same source.
 Microquasars highlight  the dramatic quasi-cyclical temporal evolution within single sources \citep{Kylafis2015}, and possibly also provide scaled down analogues for the temporal evolution of AGN. In short, all of this  highlights that we do not have as textured  an understanding of  the evolution of accretion engines as we do of stars.

Note also that despite the theoretical likelihood, observations supporting the universality of specifically disc accretion are  mostly indirect, coming from  observed spectra and luminosity considerations.  Perhaps the best evidence  in AGN  comes  from spectral analysis of the relativistic iron line profiles in Seyferts and their agreement in numerous cases
with simple relativistic flat disc models illuminated from above by coronae \citep{Risaliti2013,Reynolds2014}, although there remain  some uncertainties with respect to  how the reprocessing  and X-ray illuminating material is configured,  the interplay between continuum changes and line changes, and the associated atomic physics.  In  YSOs,  discs are observed directly \citep{Stapelfeldt2014} but  most of their emission comes in the infrared from  dust-reflected stellar light, not accretion itself \citep{Calvet2005,Kim2013}.


Mechanisms of enhanced angular momentum transport in discs can  be divided into two categories: large scale (non-local) and small scale (local) transport.
Large scale transport is an appropriate characterization when  the stress that transports the angular momentum is dominated by structures  comparable to or larger than macroscopic disc thicknesses. 
Cases of predominantly large scale transport include those dependent on  large scale magnetically mediated outflows
 \cite{Blandford1982,Konigl1989} or   large scale magnetic loops  \cite{Lynden-Bell1969}.
Small scale transport is the appropriate characterization  when the structures dominating  transport have spatial 
scales  smaller than the disc scale height.  The concept that transport can be modeled by a  turbulent viscosity  (\cite{Salpeter1964,Shakura1973}) is more  justifiable from first principles when
small scale rather than large scale structures dominate the transport.   The formalism is often used as a simplifying proxy 
even when the physical distinction between large and small scale transport is not made. 

Because of the practicality of the axisymmetric  transport model of \cite{Shakura1973} (hereafter SS73), in which the
combination of Reynolds and Maxwell radial stresses  are  replaced by a term proportional to a "turbulent viscosity", 
much research has followed in an effort to identify instabilities to drive the turbulence
that  might produce this "viscosity."  A leading candidate, and conceptual breakthrough,  has resulted from application of the  magnetorotational instability (MRI)    \citep{Velikhov1959})  to accretion discs \citep{Balbus1991,Balbus1998}.  The key insight from the latter was that in the context of Keplerian accretion discs, the linear instability criterion for the MRI requires only a radially decreasing angular velocity. In contrast, the hydrodynamic analogue that requires a radially decreasing angular momentum gradient.
An accretion disc is thus linearly unstable to the former but stable to the latter.

But after decades later of substantial and  important numerical simulation work on the MRI, there remains a significant gap between the output of simulations and incorporating them into a next generation, practical theory of accretion  beyond SS73  for use by  modelers.
(A somewhat analogous  circumstance is the relation between mixing length theory of convection to that  of the extensive numerical
simulations of convection.)   In this paper, we highlight the need to close this gap. This includes  synthesizing recent numerical results showing that transport from the MRI is not consistent with a small scale viscosity.  
In section 2 we discuss how accretion theory is normally used in model building, what is missing,  
 and how standard  SS73 accretion theory emerges from a formal mean field theory that does not include large scale transport.
 We also discuss the importance of quantifying the precision of mean field theories.
  In section 3 we discuss  lessons about  large scale transport from the observational interpretation of jets and coronae.
  In section 4 we discuss various pieces of  evidence for large scale  transport  even from shearing box simulations. 
     There we also revisit the physical interpretation of the MRI applicable to a shearing box.
In section 5 we discuss specific open questions for shearing box studies, and in 6 we discuss more fundamental
open challenges for understanding the  role  of large scale magnetic fields. We conclude in section 7.

 

\section{Standard Accretion  Theory is a Mean Field Theory Based on  Local Angular Momentum Transport}

\subsection{How Standard Disc Theory is Used}
The ultimate goal of accretion  theory is to explain and predict observations of accretes, most fundamentally, the observed flux, radiation spectra, and temporal evolution. Accretion engines  include not only  emission from the disc  but contributions from large scale outflows, extended coronae, or reprocessed emission from the central star as in the case of YSOs \cite{Calvet2005}. The ionization complexity, the mixture of material in different states of matter, and the potential role of self-gravity  make YSO accretion discs  possibly the most challenging.   A separate set of complexities for   black hole accreting disc includes the effects of  radiation pressure, 
general relativity for the most rapid rotators, and the associated fact that   outflows can be powered by magnetically mediated extraction of the rotational energy of the back hole. The latter is a separate energy reservoir from that of the accretion, although the latter would supply the magnetic field.

 \begin{figure}
  \centerline{\includegraphics[width=200px, height=150px]{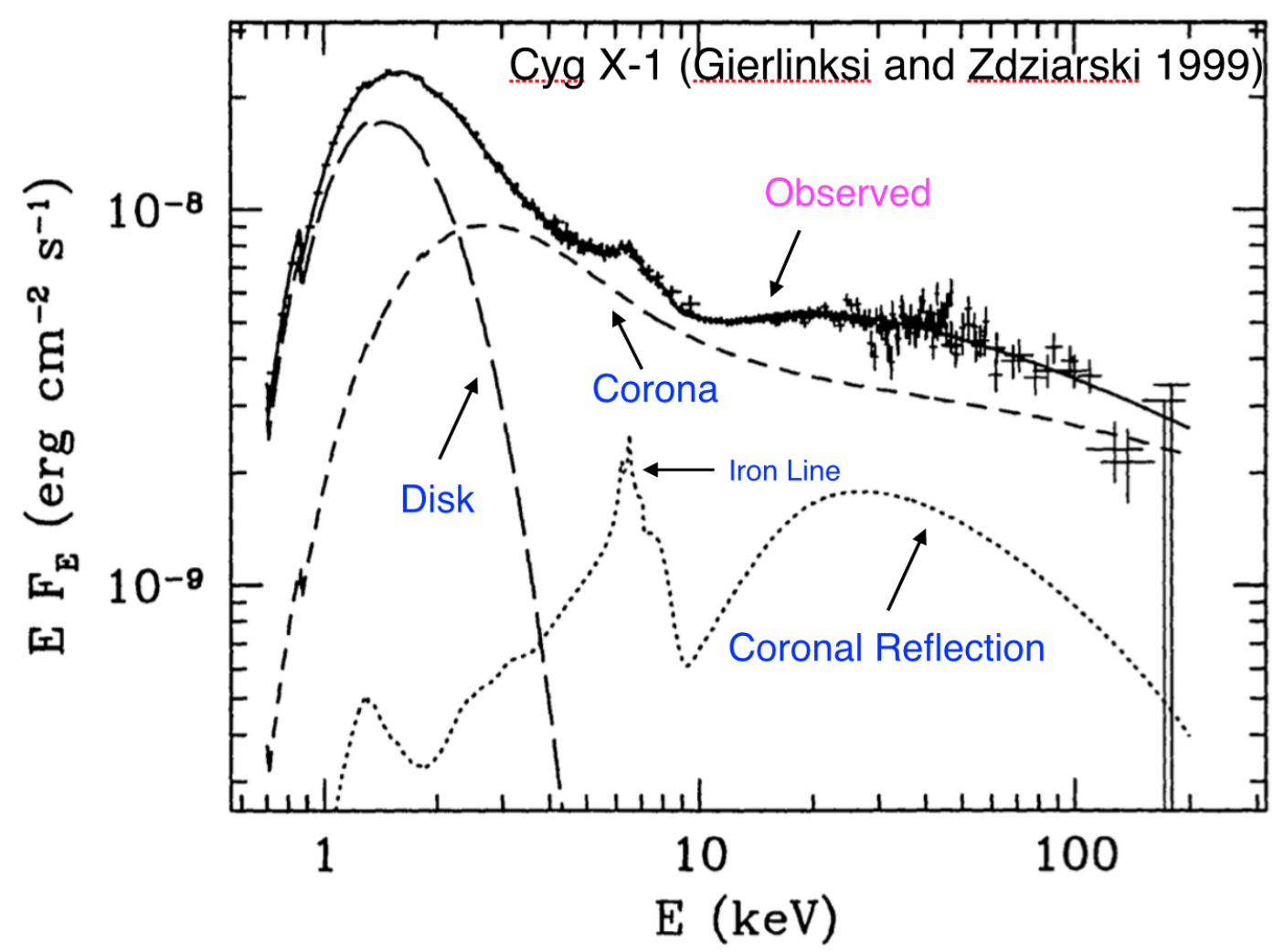}}
  \caption{Cyg X-1 flux spectrum as an example of how engine models are patched together to match observations.
Typically the optically thick continuum contribution to the disc spectrum is computed from an SS73 type model while the optically thin coronal spectrum is separately  computed from a  hot  optically thin inverse-Compton dominated plasma.  The two are then added in the relative combination needed to match the observations. The reflected spectrum is self-consistently computed given the coronal spectrum and  assumptions about the disc geometry.  What is  missing is a self-consistent  theory  that predicts, from first principles,
the relative contribution of disc, corona, and jet without having to patch together separate spectra empirically.}
\label{figure1}
\end{figure}
 
But even for an accretor without an additional  energy source from a spinning central gravitater, the accretion itself can be a source of  an optically thick disc,  optically thin corona, and outflows. Presently,  modelers commonly construct spectral models by pasting together  
 empirically weighted contributions from these ingredients  at the fractions needed to match observations.  Fig \ref{figure1} shows this for Cygnus X-1, a well known X-ray microquasar \citep{Gierlinski1999}. The three contributions needed to match the spectrum are the continuum optically thick disc emission, direct optically thin coronal emission, and disc-reprocessed  coronal emission.  The reprocessed coronal emission is determined self-consistently, but the  relative fractions of  direct coronal and disc emission are tuned to  match the observations.  A more developed theory of accretion discs  would avoid this empirical tuning.  That is,  given the engine size  (and perhaps spin) parameters along with the  accretion rate,  a developed theory of accretion  would allow us to directly predict the relative contributions of the emitting components from  and thus the overall spectra from first principles.   We would ultimately like such a theory to also predict the time evolution of these relative fractions as the accretion rate evolves.

\subsection{Standard Disc Theory Does not Distinguish Local vs. Non-Local Transport}

The standard  SS73 ``viscosity" formalism is elegantly suited for capturing the local 
radial transport of angular momentum within the disc,  but the inclusion of non-local  and vertical transport 
requires further developments. Among the important legacies of SS73 is thus the  opportunity for further work to build a more inclusive, but still practical framework.

To explain this further we emphasize that the SS73  formalism  makes sense only as a mean field theory.  This 
 follows because all dynamical quantities such as velocity and density are taken to be axisymmetric, and thus a function only of radius.
 But if angular momentum is  transported by a turbulent viscosity, axisymmetry cannot apply on arbitrarily small scales. Thus  the equations for standard disc theory must emerge from mean field equations.
To derive the theory,  dynamical quantities such as velocity, magnetic field, density, and pressure must be divided into mean and  "fluctuating" quantities, with suitably chosen averages of the latter vanishing.  While this procedure is not formally shown in SS73, a first principles approach  using with different  averaging procedures (\cite{Balbus1998,Kuncic2004}) reveals that the turbulent viscosity   emerges from a specific closure approximation to the correlated stresses associated with magnetic and velocity fluctuations.

To see  more explicitly the origin of turbulent viscosity and the limitations,  
note that the rate of change of vertical angular momentum for a magnetohydrodynamic disc  can be  obtained by multiplying the 
azimuthal $\phi$ component of  momentum density by radius in cylindrical coordinates.   This gives  \citep{Balbus1998} 
\beq
\partial_t(\rho r v_\phi) +\div \left [r\rho v_\phi \bfv-  r{\bfB B_\phi\over 4\pi} + r\left(P_{th}+ {B_r^2+B_z^2\over 8\pi} \right) {\hat {\bf e}}_\phi\right]
-\div\left[ {r\nu_V {\hat {\bf e}}_\phi \over 3}\div\bfv + r^2\nu_V\nabla{v_\phi\over r}\right]=0, 
\label{torque}
\eeq
where $\rho$ is the density, $P_{th}$ is the thermal pressure,  $r$ is the radius,  ${\bf B}=(B_r,B_\phi, B_z)$ is the magnetic field,
${\bf v}=(v_r,v_\phi,v_z)$ is the velocity, $\nu_V$ is the dynamic viscosity,
and ${\hat {\bf e}}_\phi$ is a unit vector in the $\phi$ direction. We assume radiation pressure is subdominant.

The sum of the square bracketed quantities is the total angular momentum flux.
We are  interested in the transport of angular momentum  radially and vertically but not azimuthally. 
Since the divergence of ${\hat {\bf e}}_\phi$ zero, any terms multiplying ${\hat {\bf e}}_\phi$ (including the contributions to the second and third terms from the 
$\phi$ vector components) make no contribution to the radial or vertical fluxes. The terms proportional to $\nu_V$ are 
microphysical dissipation terms and we assume here that they  are negligible  for  high Reynolds number  and magnetic Reynolds number astrophysical contexts.
Removing the $\phi$ terms and the microscopic flux terms leaves  the vertical and radial fluxes associated with macroscopic flows, namely
\beq
F_z=r\rho\left [v_\phi v_z-v_{Az} v_{A\phi} \right]
\label{vertical}
\ee
and
\beq
F_r=r\rho\left [v_\phi v_r-  v_{Ar} v_{A\phi} \right],
\label{radial}
\ee
where we have defined the Alfv\'en speed as ${\bf v}_A=(v_{Ar},v_{A\phi},v_{Az})=({B_r\over \sqrt{4\pi\rho}},{B_\pi\over \sqrt{4\pi\rho}},{B_z\over \sqrt{4\pi\rho}})$.
So far, all quantities can depend on $r,\phi, z$ and it is only after a formal
averaging  that we obtain a reduction of independent variables. 
Different  procedures involving some combination of ensemble, temporal, or spatial averages 
can be used. For compressible flows, a convenient approach is to use density-weighted time-averages and density-weighted
azimuthal averages \citep{Kuncic2004}.  
Averages of density are taken to be  time and azimuthally averaged so that splitting $\rho$ into mean and fluctuating components gives
\beq
\rho= {\overline \rho} + \rho',
\label{rho}
\ee
where $\lb \rho'\rb=\overline{\rho'}=0$ and  $\lb\rho\rb ={\overline \rho}={\int_0^{t_M} \int \rho(r,\phi,z,t+t') d\phi dt' \over 2\pi t_{M}} $, where $t_{M}$ is a long time scale
compared to turbulent turnover time scales.
Averages of velocities, Alfv\'en speeds  and their higher order  moments, are taken to be {\it density-weighted} azimuthal and time-averages and indicated by a "tilde". For example, for the velocity we have
\beq
{\tilde \bfv}\equiv\lb \bfv \rb = 
 {\int_0^T \int\rho(r,\phi,z,t+t') { \bfv}(r,\phi,z,t+t') d\phi dt'  \over 2\pi T {\overline \rho}}
\ee
with the assumption that the density and velocity fluctuations are uncorrelated, i.e $\lb \rho' \bfv'\rb =0$.
For the combination of density and two velocities  for example, this averaging procedure then gives
\beq
\lb {\rho v_r v_z}\rb  = {\overline \rho }{\tilde v}_r{\tilde v}_z+  {\overline \rho}\lb  {v'}_r  {v'}_z\rb,
\ee
where we have made the additional approximation that  triple correlations vanish to obtain the last term. Then 
$\lb \rho {v'}_r  {v'}_z\rb= {\overline \rho}\lb {v'}_r  {v'}_z\rb$ after using Eq. (\ref{rho}).

Applying the above averaging procedure to Eqs. (\ref{vertical}) and (\ref{radial})  gives
\beq 
\lb F_z\rb =r{\overline \rho}\left[ {{{\tilde v}_\phi }{{\tilde v}_z} - {{\tilde v}_{A\phi }}{{\tilde v}_{Az}} + \langle {v_\phi }'{v'_z}\rangle  - \langle {v'_{A\phi }}{v'_{Az}}\rangle } \right]
\label{vertical2}
\ee
and
\beq
\lb F_r\rb =r{\overline \rho}\left[ {{{\tilde v}_\phi }{{\tilde v}_r} - {{\tilde v}_{A\phi }}{{\tilde v}_{Ar}} + \langle {v_\phi }'{v'_r}\rangle  - \langle {v'_{A\phi }}{v'_{Ar}}\rangle } \right].
\label{radial2}
\ee
These mean fluxes are functions of $r$, $z$, and coarse grained in $t$. 

To further simplify  without completely eliminating the $z$ dependence, we  integrate separately over hemispheres.
This additional coarse graining is  indicated by the subscript ``$\overline \rho$". 
Thus $\lb{\tilde v}_z\rb_{\overline \rho}\equiv {\int_0^{H/ 2} {\overline \rho} {\tilde v}_z dz\over \Sigma/2 }$, where  $\Sigma /2= \int_0^{H/2} \overline \rho dz$ and
$H$ is the disc thickness defined by a density scale height. We also assume that  the three quantities ${\tilde v}_\phi$,  ${\tilde v}_r$, and ${\tilde v}_{A\phi}$ are independent of $z$.  The results of the $z$ integration of Eqs. (\ref{vertical2}) and (\ref{radial2}) are then
\beq
\lb F_z\rb_{\pm} =\frac{{\Sigma r}}{{2}}\left[ {{{\tilde v}_\phi }{{\tilde v}_{z\pm,\overline \rho}} - {{\tilde v}_{A\phi,}}{{\tilde v}_{Az\pm,{\overline \rho} }}+ \langle {v'_\phi }{v'_z}\rangle_{\pm,\overline \rho}  - \langle {v'_{A\phi }}{v'_{Az{\pm,\overline \rho}}}\rangle_{\pm,\overline \rho} }\right]
\label{vertical3}
\ee
and
\beq
\lb F_r\rb_{\pm} =\frac{{\Sigma r}}{{2}}\left[ {{{\tilde v}_\phi }{{\tilde v}_r} - {{\tilde v}_{A\phi }}{{\tilde v}_{Ar{\pm,\overline \rho}}} + \langle {v'_r}{v'_\phi }\rangle_{\pm,\overline \rho}  - \langle {v'_{Ar}}{v'_{A\phi }}\rangle_{\pm,\overline \rho} } \right],
\label{radial3}
\ee
where the $\pm$ subscripts indicate the values in the ``+" (upper) and ``-" (lower) hemispheres, but there is otherwise no  $z$ dependence left.
In each hemisphere, the quantities remain  functions of radius and coarse grained time.  The stresses presented in this form
offer a unifying focal  point for a variety of  generalizations of SS73, and keeping the hemispheres averaged separately is relevant for
tracking any pseudo scalars (such as  helicities) or fluxes into coronae that might play a dynamical role.

To see SS73 emerge from these equations, we  
take  ${\tilde v}_z={\tilde v}_{Az}={\tilde v}_{Ar}=0$, and  ignore vertical stresses from fluctuations.
Then, only the radial flux from Eq. (\ref{vertical3}) matters.  If we further   assume the $\pm$ contributions are equal (symmetric across the mid plane)
then
\beq
\lb F_r\rb = \lb F_r\rb_{+}+\lb F_r\rb_{-}
 ={{\Sigma r}}\left[ {{{\tilde v}_\phi }{{\tilde v}_r} +{{T}_{r\phi}}}\right],
\label{radial33}
\ee
where ${{T}_{r\phi}}$ is now the   stress combination from fluctuations given by 
\begin{equation}
{{ T}_{r\phi}} = \langle {v'_r}{v'_\phi}\rangle  - \langle {v'_{Ar}}{v'_{A\phi}}\rangle  \simeq  - {\nu _{T}}r{\partial _r}{{ \Omega}}\sim q  \alpha_{ss} c_s^2,
\label{usual}
\end{equation}
where the  penultimate similarity  (Lynden-Bell \& Pringle 1974)  represents the closure that converts a sum of second order fluctuation correlations into a term linear
in  the  angular velocity shear. The latter similarity follows from setting $\nu_{T}\equiv\alpha_{ss} c_s H$, where  $\alpha_{ss}$ is the dimensionless SS73 ``viscosity" coefficient, 
along with $\Omega \simeq c_s/ H$ in hydrostatic equilibrium, and  $\Omega\propto r^{-q}$, where $q=3/2$ for Keplerian flow.
 Although SS73 warned that $\alpha_{ss}$ could be a function of radius, it has become common practice to employ the simplest model in which  $\alpha_{ss}$ is a constant.

We can now recover a standard  form of the accretion rate balance equation in the steady-state.
   Applying  Gauss' theorem  to Eq.  (\ref{torque})  in  cylindrical coordinates with only a radial derivative and integrating the flux divergence when vertical stresses are ignored,  implies that $r\lb  F_r \rb$  is constant.  If we  further assume that   stresses from fluctuations vanish at the inner boundary $r=r_0$,   then  using $r\lb F_r \rb|_{r=r_0}=r\lb  F_r \rb$    in Eq. (\ref{radial33})   along with
  ${\dot M}=-2\pi\Sigma r {\tilde v}_r $ for the steady-state accretion rate, we obtain
\beq
-{{\dot M}r_0^2\Omega_0\over 2\pi }=-{{\dot M}r^2\Omega\over 2\pi}+r^2\Sigma T_{r\phi}. 
\label{stress1}
\ee
Thus
\beq
 T_{r\phi} = {{\dot M}\Omega \over 2\pi \Sigma} \left(1- {r_0^{2-q}\over r^{2-q}}\right) 
\label{stress2}
\ee
for   ${\tilde v}_\phi=\Omega r$,  and  $\Omega=\Omega_0(r/r_0)^{-q}$.
 A positive stress implies  positive
accretion for $q <2$.   For $q>2$,  this equation alone would suggest excretion  but  for this regime epicyclic frequencies become imaginary and 
 the orbits and disc are  unstable, violating the validity of the steady-state assumed above, and invalidating the excretion interpretation.
This is also why  the shearing box model produces runaway mean velocities for $q>2$ and is  unsuitable  for that regime \citep{Nauman2014}.

The above derivation highlights  that a physically consistent derivation of the  standard $\alpha_{ss}$ viscous disc accretion model 
from the MHD equations requires the following:
  {\it (i)} formal averaging that   leaves the mean field quantities dependent only on radius;  
  {\it (ii)} dropping  vertical transport terms associated with  large scale fields; {\it (iii)} dropping vertical transport terms associated  small scale MHD fluctuations; {\it (iv)} dropping  radial transport terms associated with large scale magnetic fields; {\it (v)} including the radial transport from fluctuations by replacing 
   the sum of their stress contributions by a viscous term linear in the mean angular velocity gradient. 
   There is a  significant  opportunity to improve  this standard theory  by  relaxing even one of the above assumptions.
 The  importance of large scale fields, as evidenced by jets
  or coronae,  often invalidates assumption at least assumption {\it (ii)} above. In addition,  vertical stresses associated with fluctuations can be smaller  than the radial stresses  by a factor $H/r$ but still contribute equally to the torque when both radial and vertical torques are present, violating assumption {\it(iii)}.
Radial transport by large scale loops is excluded by assumption {\it (iv)}, which may also also be a significant restriction (see also \cite{Rudiger87}).
Note that even if stresses from large scale structures might be written as proportional to $c_s^2$ by analogy to that of viscous transport, the dimensionless coefficient can be different for that contribution,  corresponding to  large scale structures which fuel coronae and outflows compared to local viscous transport which supplies
heating within the disc.

The standard SS73 closure modeling the transport as purely viscous is a simple and elegant, but
there have been early hints  that it may be  misleading when applied to MRI driven turbulence \citep{Balbus1998}. 
There have been specific attempts to improve the closure \citep{Ogilvie2003,Pessah2006},  and more recent direct evidence  
that the MRI stress does not agree with the SS73 formulation \citep{Pessah2008,Nauman2015}. 
As discussed further below, our interpretation of  numerical evidence further suggests  that the  transport  from   the MRI is actually dominated by large scales.

How does previous work on accretion disc theory connect to relaxing  assumptions $(i-v)$?
\cite{Blandford1982} and \cite{Konigl1989} exemplify  models for which the primary angular
momentum transport occurs via large scale magnetic fields.  Such models implicitly  include 
{\it (i), (ii)} and {\it (iv)}, but not {\it (iii)} and {\it (v)}.
\cite{Blandford1999}  and \cite{Quataert1999} are examples of  practical models  which include mass loss
to reduce disc luminosity,  and thus implicitly the role of vertical transport, but the associated stresses are  not formally derived from a closure nor are the outflows emergent from equations that include large scale magnetic field dynamics.  The  models are  rooted in the viscous formalism of assumption {\it (v)} with the additional assumption that in collisionless plasmas,  the dissipation can may heat primarily ions not electrons, and the coupling of the two being limited primarily by Coulomb collisions.  Understanding the physics of this dissipation 
\citep{Blackman1999}
and energy transfer  is itself  an important area of active research at the small scale, collisionless frontier of  accretion theory \citep{Quataert2014}, but is outside of the scope of the present paper.

 \cite{Ruediger1993b,Campbell1999,Campbell2000} developed models for accretion that transfer angular momentum via  large scale 
 $\alpha-\Omega$ dynamo-produced fields, and showed that this contribution to transport  can be comparable to that from turbulent transport of the SS73 type. \cite{Campbell2003} solved for the vertical structure and radial structure  in such  discs.  Taken together, these papers  maintain assumption {\it v} above but relax the others. They should  be revisited with an eye toward reducing the complexity and to incorporate new developments  from large scale dynamo theory (see sections 4.2 and 4.3).  
 A separate important step towards a first principles improvement to mean field disc theory  is \cite{Kuncic2007} which adopts  {\it i, ii} and {\it iv} and {\it v}   but drops {\it iii} by including  some vertical transport.  
 \cite{Campbell1999,Campbell2000,Campbell2003} did not compute the  observed spectral influence of the vertical transport and 
 \cite{Kuncic2004,Kuncic2007}  partially did but only by removing it from the contribution to the disc  spectra. 
Overall, there remains much opportunity for further work.

 
\subsection{Averaging and Precision of Disc Theory}

That standard $\alpha_{ss}$  accretion disc theory is a mean field theory 
means that  the precision of the predictive power 
is limited in spatial, temporal, and  spectral resolution.   Without quantifying this precision and recognizing that it is finite, 
a disagreement between theory and observed data can  be misinterpreted to imply the need for new physics rather 
a consequence  of statistical fluctuations within the predicted precision error.
 
To obtain to the standard link between accretion and stress in the previous section,  we considered a time-average and a $z$-integration without  averaging in $r$. The longer the duration over which the time-average is taken, the lower the predicted error in  precision. 
In comparing theory with data, it is important to know the  time scales over which the data are binned to be sure  that appropriate theoretical averages are being used to  compare to with appropriately averaged data.
The smaller the scales of turbulent fluctuations compared to the smoothing scales, 
the more agreement (and thus freedom in choice) there  is between  temporal, ensemble, and quasi-local spatial averages.
 However, if fluctuations  are  sufficiently large, then there may be a mismatch  between predictions of the  theory and observations taken over  very short integration times even if the theory would correctly explain the data averaged over much longer times.
The case of YSOs is particularly noteworthy because data against which theory is often compared are integrated for time scales much smaller than an orbit time.  
 Some work on this mismatch between the statistically expected deviations from the predicted
 mean field theory has been done \citep{Blackman1998,Blackman2010} but more is needed to address the non-local contribution
 of emission to a given wavelength.

A related  subtlety is that although variability and deviations from standard predictions of axisymmetric theory may  arise from systematic 
phenomena such as warps \citep{Hartnoll2000,Tremaine2014}, when the theory itself uses a viscous model for transport, self-consistency requires an assessment that the predicted structures are  large compared  to scales of the statistical fluctuations over which the mean field theory is averaged.

\subsection{Local Radial Transport is Insufficient: Vertical and Nonlocal  Transport Must be Included}

Although most accretion disc modeling  presumes  a viscosity transport model and a constant $\alpha_{ss}$,  
 observations and simulations  challenge  this minimalist paradigm.
Evidence for its insufficiency emerges from: (1) observations  indicating  non-local radial and vertical transport via jets, outflows, and coronae and the important role of large scale fields in  plausible jet models.
(2)  inadequacy  of total stress magnitudes derived from local shearing box MRI simulations if taken at face value; (3)  dominance of  large scale stresses in local and global simulations;  (4)  contradiction between transport computed from simulations 
 and the  dependence on shear predicted from the constant $\alpha_{ss}$ viscosity model.
 (5)   evidence that  $\alpha_{ss}$  depends on radius in some global simulations \cite{Penna2013}.
We discuss points 1-4 further below.

 \section{Lessons From Observations of Jets, Coronae, and Large scale  structures}
 
Jets and coronae are conspicuous components of accretion engines and highlight the importance of vertical transport. 
Seyfert AGN observations   reveal  $ \ge 30$\% of bolometric emission coming from corona-produced X-rays \citep{Mushotzky1993} 
and in some radio sources and blazars,  jets even have more mechanical luminosity 
than the underlying discs alone can provide \citep{Ghisellini2014} (requiring input from the central rotator).
These circumstances indicate the prevalence of large scale modes of transport. 
In sub-Eddington sources, Large scale magnetic fields provide  natural   angular momentum transport  agents  \citep{Blandford1982,Konigl1989,Field1993,Blackman2001,Lynden-Bell2006,Pudritz2012,Penna2013}.  In young stars,  pre-planetary nebulae, microquasars, and active galactic nuclei, jets typically  have too much collimated momenta to be driven by  mechanisms that do not involve  large scale magnetic fields \citep{Pudritz2012,Blackman2014}. In the jets of AGN, Faraday rotation from ordered helical magnetic fields is directly observed \citep{Taylor1993,Asada2008,Gabuzda2012}.  

Magnetized outflows  anchored in  accretion engines  can extract  angular momentum without expelling much mass,  allowing the remaining disc material to accrete.  
Jet formation may depend on role large scale fields  also play in coronae:  if discs are  turbulent, then coronae likely emerge from disc  magnetic ejecta whose structures are large enough to escape  turbulent shredding on a buoyant rise time \citep{Blackman2009}.   Once these structures arrive to the coronae, they can further open up by dynamical relaxation  to form large scale fields that  facilitate jets.   In many objects,  broader winds, in addition to jets, may also provide a loss of
 angular momentum.  The common presence   of corona, jets, and outflows  raises a generic question of whether 
accretion may in fact {\it require} outflows. This  highlights  a substantial  deviation from  solely local radial transport of angular momentum.

Outflows  and coronae  can in principle be included in  extensions of  disc models with zero-stress internal boundaries and steady accretion,  but real discs also have more varied boundary conditions that lead to  additional large scale features. For example, the magnetospheric coupling between disc and stellar field   
  has been long studied in the context of compact objects \citep{Ghosh1978,Perna2006} and YSOs   \citep{Matt2005,Romanova2012,Lii2014,Lai2014}.   
     Large scale vortices,  either  hydrodynamically    \citep{Lovelace1999,Li2000,Colgate2003,Klahr2003,Barranco2005}  or  magnetically induced  \citep{Tagger1999,Varniere2002}, also provide  modes of non-local transport.  
   Self-gravity can also produce large scale features  
     such as spiral arms that  induce transport and encompass another set of processes
 \citep{Gammie1996,Gammie1998,Vorobyov2007,Zhu2009}.
 In YSOs, large scale disc structures are now observed with ALMA \citep{vanderMarel2013,Perez2014}. 
 
 \cite{King2007,Kotko2012} point out that   Dwarf Novae and AGN   
  require $0.1 \lesssim \alpha_{ss}\beta \lesssim 0.4$ in the SS73 viscous paradigm, which is   a factor of 10 larger than what the   "best converged" modern shearing box MRI simulations produce e.g. \citep{Davis2010}.
 Because of the  limitations of the  theory and the simulations, lingering questions about  convergence \citep{Bodo2014}, and the physical applicability of
 what is learned form the shearing box  can be debated, but the factor of 10  discrepancy may be indicative of 
 transport occurring on much larger scales than what a shearing box is able to capture.
 
\section{Lessons from  Shearing Boxes}

The widely used  local shearing box is a very powerful tool  \citep{Balbus1998} to study aspects of transport and the associated nonlinear MHD  even though it fundamentally differs  from  a  local section of a real disc. 
The shear periodic radial boundaries reduce the accessible phase compared to a real system and there is a symmetry between outward and inward radial directions that is broken for a real disc because the latter has a finite radius of curvature. The boxes also can lead to an 
 an injection of  Poynting flux due to the non-periodicity of the azimuthal velocity  \citep{Hubbard2014}. Shearing boxes to date also do not include vertical shear \citep{Pessah2014}.

Since the radius of curvature is infinite for a shearing box, the angular momentum is ill-defined  and there is no finite divergence of the angular momentum fluxes  and thus no torque.  What we learn about transport from shearing boxes comes from computing the stresses.
 As mentioned above,  the basic question of convergence  of stresses is still under debate \citep{Davis2010,Bodo2014}. 
 Convergence of stresses is not yet achieved for global simulations  \citep{Hawley2011} and more subtle questions of convergence of spectral shapes have yet to be fully vetted \citep{Nauman2014}.
Before discussing lessons about large scale transport learned form these stresses, we 
we first discuss the MRI a bit further in the context of shearing boxes to clarify the physical interpretation of the instability therein.
 
\subsection{ Understanding  the MRI in shearing boxes as a transfer of linear 
momentum}

A widely used physical explanation  for the linear MRI  appeals to two masses tethered by a spring 
separated by an initial radial displacement in a background rotation profile of outwardly increasing angular momentum \citep{Balbus1998}.
For a weak enough spring,  the inner mass loses angular momentum to the outer mass and will continue to move toward loci
of lower  background angular momentum, namely, inward.  Complementarily, the outer mass  moves toward regions of higher background angular momenta, namely outward.
By considering the equations for each mass separately and taking their difference, the expression for the separation between the masses can be obtained
with  exponential growth being the signature of the instability. A similar approach can be used to argue that  a flow with an outwardly decreasing background angular momentum gradient should be stable to the MRI. Such stability is consistent with recent generalizations \cite{Shakura2014}.  

However, because the Cartesian shearing box has an infinite radius of curvature  the angular momentum is ill-defined.
But the  instability is still robust therein because it manifests in the linear momentum equations.  The angular momentum transport is a secondary consequence  that arises for a finite radius of curvature.  Although these points  follow simply  from the known equations,  the standard physical explanation of the MRI given in the previous paragraph with angular momentum commonly presented without clarification for a shearing box. To  make the  clarification clear, we  discuss  the tethered spring analogue in more detail below.  
 
To converting apply the viscous closure Eq. (\ref{usual})  to that of a  shearing box, 
we shift to the local Hill frame Cartesian coordinates for a  disc  co-rotating with  local  angular speed $\Omega_0= \Omega(r=r_0)$,  assumed to represent the range   $r_0-x \le r \le  r_0+x$,  
with  $x <<r_0$.  Ignoring curvature terms, the  
 mean $y$-velocity in the box  equals
\beq 
 {\tilde v}_\phi-\Omega_0 r_0=r (\Omega-\Omega_0) \simeq x r\partial_r \Omega|_{r=r_0}= -x q \Omega
 \label{boxv}
 \ee
  to lowest order in $x$ so that 
  \beq
   \lb{{T}_{r\phi}}\rb=\alpha c_s H q \Omega_0
   \label{boxv2}
   \ee
    in place of  Eq. (\ref{usual}).
From the right side of Eq. (\ref{boxv}) we see that the center line of the box  $x=0$  represents  the co-rotation radius whose mean azimuthal ($y$-direction) velocity vanishes.

In the absence of pressure gradients, the Hill frame equations capturing the  MRI instability  \citep{Balbus2003} in the presence of a vertical field of strength $B_z$,  subjected to displacement  $\boldsymbol{ \xi}=(x(t) e^{ikz},y(t)e^{ikz})$ about the vertical line $(x=0, y=0)$  can be written 
\beq
{\ddot x} -2\Omega {\dot y} = -(K_A- T) x,
\label{hill1}
\eeq
and
\beq
{\ddot y} +2\Omega {\dot x} =  -K_A y,
\label{hill2}\eeq
where 
$T= -r d\Omega^2/dr=2 q \Omega^2$  ($q=3/2$ for Keplerian)
 is  the coefficient of the  tidal force per unit mass in the  Hill frame; 
   $K_A=(k{\tilde v}_{zA})^2$, where ${\tilde v}_{zA}$ is the Alfv\'en speed associated with the vertical field; and the second terms on the left sides come from the Coriolis force.   The appearance of $K_A$ arises as a consequence of  magnetic tension  force  because 
 the linear fluctuation in the field $\delta \bfB$ satisfies $\partial_t \delta \bfB\sim  ikB \delta \bfv$ from the induction equation,  so that  $ \delta \bfB\sim ikB{\delta{\bf v} t}\sim  ikB\boldsymbol{\xi}$ upon time integrating.  The curvature force appearing on the right side of the momentum equation  is then  ${\bfB\cdot \nabla\over 4\pi \rho}\delta  \bfB\sim -K_A \boldsymbol{\xi}$ with the magnetic pressure  having been absorbed into the total pressure, assumed to have a vanishing gradient.
 
Since the $z$ dependence  dropped out of Eqs. (\ref{hill1}) and (\ref{hill2}), the relevant motion takes place in the $x,y$ plane.
The equations can then be interpreted to describe the evolution of the position of a small parcel of fluid at position $\boldsymbol{\xi}$ tethered by spring
to a fixed point at the origin $x,y=0$. Given that $K_A$ and $T$ are constants, the right hand side of the above equations then represents
 two directional components of a spring force per unit mass with distinct  spring constants  $K_A-T$ and $K_A $ respectively.  
The spring constant in Eqn. (\ref{hill2}) has a positive sign, such that  positive  displacement strengthens the force opposing the displacement.  However,  if  $T> K_A$ then Eq. (\ref{hill1})  would have a  negative spring constant, so the associated  force would exacerbate the separation of the fluid parcel from the origin as the separation increases in the $x$ direction.
 When interpreted as a runaway displacement of the fluid parcel from its initial position,  instability  requires $T>K_A$ and stability requires $T\le K_A$ as found by solving the two equations.  (Note that for $q <0$,  the azimuthal velocity increases outward. Then 
$T<0$,  and the effective spring constant in Eq. \ref{hill1} 
would be positive, implying stable oscillations.)
 
Eqs. (\ref{hill1}) and (\ref{hill2})  equation can also be adjusted  to describe the displacement  of two equal density fluid parcels mutually  attached by a   spring.
To do so, we must construct two sets of equations similar to   Eqs. (\ref{hill1}) and (\ref{hill2}), one for each of the two masses and then subtract them to get the
equations of evolution of their separation.   However note that while the  tidal force has the same form as in the above equations, the force from the $K_A$ 
terms depends on the separation between the two masses. The equations for each of the two parcels thus take the form
\beq
{\ddot x}_i -2\Omega {\dot y}_i = Tx_i- K_A(x_i-x_j),
\label{hill1a}
\eeq
and
\beq
{\ddot y}_i+2\Omega {\dot x}_i =  -K_A (y_i -y_j),
\label{hill2b}\eeq
where $i=1,j=2$  represents the equations of motion for parcel 1 and $i=2, j=1$  represents the equations for parcel 2.
Subtracting the respective $x$ and $y$ equations for parcel 2 from those of parcel 1, and letting $(\zeta_x,\zeta_y )= (x_1-x_2, y_1-y_2)$ 
gives
\beq
{\ddot \zeta}_x-2\Omega {\dot \zeta}_y= -(2K_A- T)\zeta_x
\label{hill1aa}
\eeq
and
\beq
{\ddot \zeta}_y+2\Omega {\dot \zeta}_x=  -2K_A\zeta_y.
\label{hill2bb}
\eeq
Instability as interpreted as  separation of the two connected masses requires $T> 2K_A$.
Neither of the above  explanations of the instability  criteria require explicit mention of  angular momentum.
The instability is evident in the linear momentum equation  and the stress that it produces in the shearing box limit represents a transfer of linear momentum. 
For a finite radius of curvature, angular momentum transport then becomes a physical consequence.

\subsubsection{Laboratory experiment to test the spring-tethered masses model of the MRI?}

We suggest that  existing experimental configurations set up to ultimately study the MRI using liquid metals (e.g. \cite{Ji2011}) could in fact be useful for also setting up a tethered mass MRI analogue experiment. Two  light masses tethered by a weak spring
could be placed in a Taylor-Couette  flow tank using even water, and their  motion tracked.  Their evolution would provide data for  comparison with linear theory and for comparison with the results  expected for a system with  a central gravitational potential.  
Drag forces and boundary layers around the embedded masses
may provide complications but such an experiment  warrants further investigation as it is 
 conceptually  relevant and likely inexpensive.

\subsection{Stress Spectra from Shearing Boxes: The Dominance of Large  Scales} 

\subsubsection{Stress Spectra From Fluctuations}
The simplifications and limitations of the shearing box are significant compared to real discs, 
but, like the triply periodic box commonly used in forced MHD turbulence studies,  the shearing box should be thought of as its own well-defined computational physics experiment.   If we can understand how fields and flows
evolve in such suitably chosen minimalist systems, we hopefully gain insight into  key elements of  real systems.

How well are the basic assumptions of the SS73 paradigm  confirmed or refuted by shearing box studies? This is  germane given that stress magnitudes from shearing box simulations are  widely quoted as inputs for spectral models.
Shearing boxes were designed  with the intuition that angular momentum transport is a primarily local, radial transport phenomena.
  In this context we can ask  {\it (i)} what scales dominate the stresses? {\it (ii)}  do the radial stresses depend linearly on shear as predicted by the SS73 framework with constant $\alpha_{ss}$?   As we will see below, the answers are: 1) large 2) no.

\begin{figure}
  \centerline{\includegraphics[width=300px, height=225px]{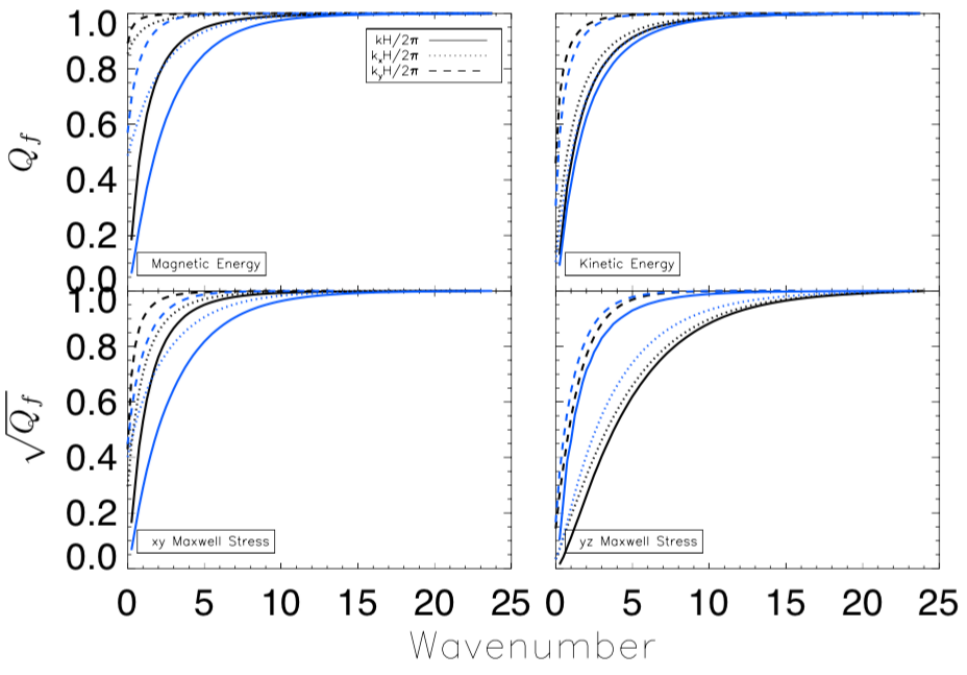}}
   \caption{Modified from \cite{Nauman2014} (to include $(0,k_y)$ and $(k_x,0)$ modes):   the fractional spectral power for  several  quantities as indicated in a Keplerian shearing box simulation (see text for parameters) vs. a dimensionless measure of the wave number. 
   Black curves are for corona $|z|\ge 2H$ and blue curves are for disc $|z|<2H$ and the legend shown applies to both color sets of curves.
   The wave numbers are of the form $k=2\pi n/L$ where $L=4H$ is the domain size in the $x$ or $y$ direction
   and $n$ ranges from 1 to 96. The x-axes  are  $kH/2\pi =n/4$.   
     The total stresses are dominated by the Maxwell stresses  and so lessons learned from their plots also apply for the total.  
     Both coronae and disc show that the fractional power for all quantities is well over 50\% already for the first few wave modes.
    Fractional power is calculated using equation \ref{Qs} using $|f({\bf k})|^2 = \left\{ |B({\bf k})|^2, |{\sqrt \rho} v({\bf k})|^2, |B_x B_y ({\bf k})|^2, |B_y B_z ({\bf k})|^2\right\}$ for magnetic energy, kinetic energy, xy Maxwell stress, yz Maxwell stress respectively. For energy $Q_f$ gives the fractional power, but for the stresses it gives fractional power of stress squared, so we plotted $\sqrt{Q_f}$ in those cases. The fractional power distribution in $k_y$ is significantly different from that of $k_x$ and $k$ for all quantities, highlighting anisotropy. 
}
\label{fig2}
\end{figure}

We  first determine  the scale dependence of  stresses from from  stratified  shearing box simulations 
by computing the  $k$-dependent cumulative stress from fluctuations  \citep{Nauman2014}.  This is computed by integrating from the minimum wavenumber up to $k$  and determining how large $k$ must be in the integral to capture most of the stress in the box.   We  have used the \textsc{ATHENA} code and solved the vertically stratified MHD equations without explicit diffusivities, for  an isothermal equation of state. The boundary conditions are  shear periodic in the radial $x$ direction, periodic in the azimuthal $y$ and vertical $z$ directions 
 (with gravity thus taken to vanish at   vertical boundaries). 
 For the results  shown in Fig \ref{fig2},  we employed a box domain size of $4H \times 4H \times 8H$ where $H$ is the density scale height, with a resolution of 96 zones per $H$ and
 an initial plasma $\beta$ (thermal to magnetic pressure ratio) equal to 100 everywhere and an initially toroidal mean field which decreases in strength vertically like the density. 

We obtain the cumulative stresses  from fluctuations as a function of wave number by first
computing 2-D Fourier transforms $f({\bf k},z)=f({k_x,k_y},z)$  of a physical quantity $F({\bf x})$ (which was either the magnetic field,   off-diagonal magnetic stress,  momentum density, and off-diagonal Reynolds stress) in variables $x$ and $y$. 
  For each quantity we then computed three  different integrals of the quadratic norm:
   (i) circle average:   $ |f(k,z)|^2 = \int |f({\bf k}')|^2 \delta (|{\bf k}'| - k) d{\bf k}'$, where $|{\bf k}| = \sqrt{k_x^2 + k_y^2}$, 
(ii) y-integrated:  $|f(k_x,z)|^2 = \int |f({\bf k})|^2 dk_y$, 
(iii) x-integated  $|f(k_y,z)|^2 = \int |f({\bf k})|^2 dk_x$.
To eliminate $z$, we then  averaged the resulting 1-D spectra over $z$ ranges specified by either "disc" or "corona" as determined by the height above which  the saturated $\beta$ is above or less than unity respectively. 
The cumulative stress and energy spectra  then  represent the integral of a given spectrum up to
wave number $k$ divided by the integral over the full range of $k$, that is:
\beq
\begin{array}{l}
{Q_f}(k) = \frac{{\int_{1}^k {|f(k'){|^2}2\pi k'dk'} }}{{\int_{{1}}^{{k_{max}}} {|f(k'){|^2}2\pi k'dk'} }}\\
{Q_f}({k_x}) = \frac{{\int_{{0}}^{{k_y}} {|f({\bf{k}}){|^2}d{k_y}/\int_{}^{} {d{k_y}} } }}{{\int_{{0}}^{{k_{y,max}}} {|f({\bf{k}}){|^2}d{k_y}/\int_{}^{} {d{k_y}} } }}\\
{Q_f}({k_y}) = \frac{{\int_{0}^{{k_x}} {|f({\bf{k}}){|^2}d{k_x}/\int_{}^{} {d{k_x}} } }}{{\int_{{0}}^{{k_{x,max}}} {|f({\bf{k}}){|^2}d{k_x}/\int_{}^{} {d{k_x}} } }}.
\end{array}
\label{Qs}
\ee
The  stresses computed as above do not include contributions for which  both $k_y=0$ and $k_x=0$, though each can separately
vanish. We discuss the  $k_y=k_x=0$ contributions  in subsection 4.2.2.

Fig. \ref{fig2}  shows that, for both coronae and disc regions, more than half of the radial and vertical stresses are accounted for by contributions from the first few wave numbers. 
We found that this conclusion is independent of the box size for domain sizes studied.  The cumulative stresses  are dominated by correlations from large scales of the box. Even though  stresses are nearly constant as the box sizes and resolutions change (a 
necessary condition for convergence), the  upper integration bound on the wave number  needed to capture  50\% of the cumulative stress does not  increase as the box domain size increases.  This  suggests that the local simulations are insufficient to capture significant large scale contributions to the transport.  That in turn highlights a limitation of the shearing box; the box size cannot be increased indefinitely without having to incorporate curvature terms or  vertical shear \citep{Pessah2014} to 
ensure relevance when scaled to a real system. That  shearing boxes suggest non-local transport in the sense described  is further supported by measurements from global simulations  of energy  spectra \citep{Flock2012,Suzuki2014} and stress spectra \citep{Beckwith2011,Sorathia2012,Parkin2013}.  

While  energy spectra and stresses  for both  disc  and coronae  (respectively distinguished by plasma $\beta$ greater than or less than unity) in cases with net flux and without explicit dissipation are dominated by large scales in both local and global simulations  and the magnetic energy spectra generally do not show a turnover at large scales \citep{Nauman2014},   
the  zero net flux, unstratified simulations of  \citep{Fromang2010} with  explicit dissipation do seem to show a turnover in the magnetic and kinetic energy spectra. The wave number of the magnetic peak seems to  increase with increasing magnetic Reynolds number.
 A concern about  magnitude of stress in previous unstratified  zero flux simulations however is that they did not seem to converge \cite{Pessah2007}, 
 although \cite{Fromang2010}  argues for convergence.  Controversy still lingers   over the  question of convergence of the stress magnitude,  even for stratified simulations \citep{Davis2010,Bodo2014}.
Stress spectra, in addition the stress magnitude, should be assessed in  convergence studies.


 \subsubsection{Comparing Stresses From Mean Fields to those From Fluctuations}

The analysis of the previous section does not include contributions to the stress from $k_x=k_y=0$ modes at fixed $z$.
We  refer to these as stresses from the mean field.  Here we compare the contribution from  their stresses to those from the fluctuations.
As the measure of stresses  and energies from mean magnetic fields we compute $T_{M,ij}(z)=\lb B_i \rb \lb B_j \rb$ where the brackets indicate $x,y$ averages,  with the  product being left as a function of $z$.  We then average the  product over $z$ and orbit times $101<t<220$ to obtain the total contribution  $T_{M,ij}$.  For the measure of stress from  fluctuations, we  first  compute the contribution from the total field $T_{ij}(z)\equiv \lb B_i  B_j \rb$,  average  over $z$ and $101<t<220$ as above to obtain  ${ T}_{ij}$, and  then subtract 
the total contribution from the mean fields to obtain $ { T_F}_{ij}= { T}_{ij}-T_{M,ij}$.

Using the above procedure for  a $q=3/2$ case from a run with  32 zones/$H$ and domain size $H\times 2H \times 4H$ \citep{Nauman2015},  we find that the ratio of the stresses from the mean field to that of the total  to be  $ T_{M,xy}/T_{xy} \simeq 0.24$ and that the  ratio of  magnetic energy from the mean field that of the total  to be  $0.47$.
These indicate a  significant contribution from the mean field, bolstering  the overall message that large
scale fields  and transport are  important.  The  simulation from which we derived this result has no extended corona and so the stress ratio  likely underestimates the  importance of mean fields:  Fig. 3. shows that already at the base of the corona the mean stresses are dominant.

 \begin{figure}
  %
  \centerline{\includegraphics[width=300px, height=225px]{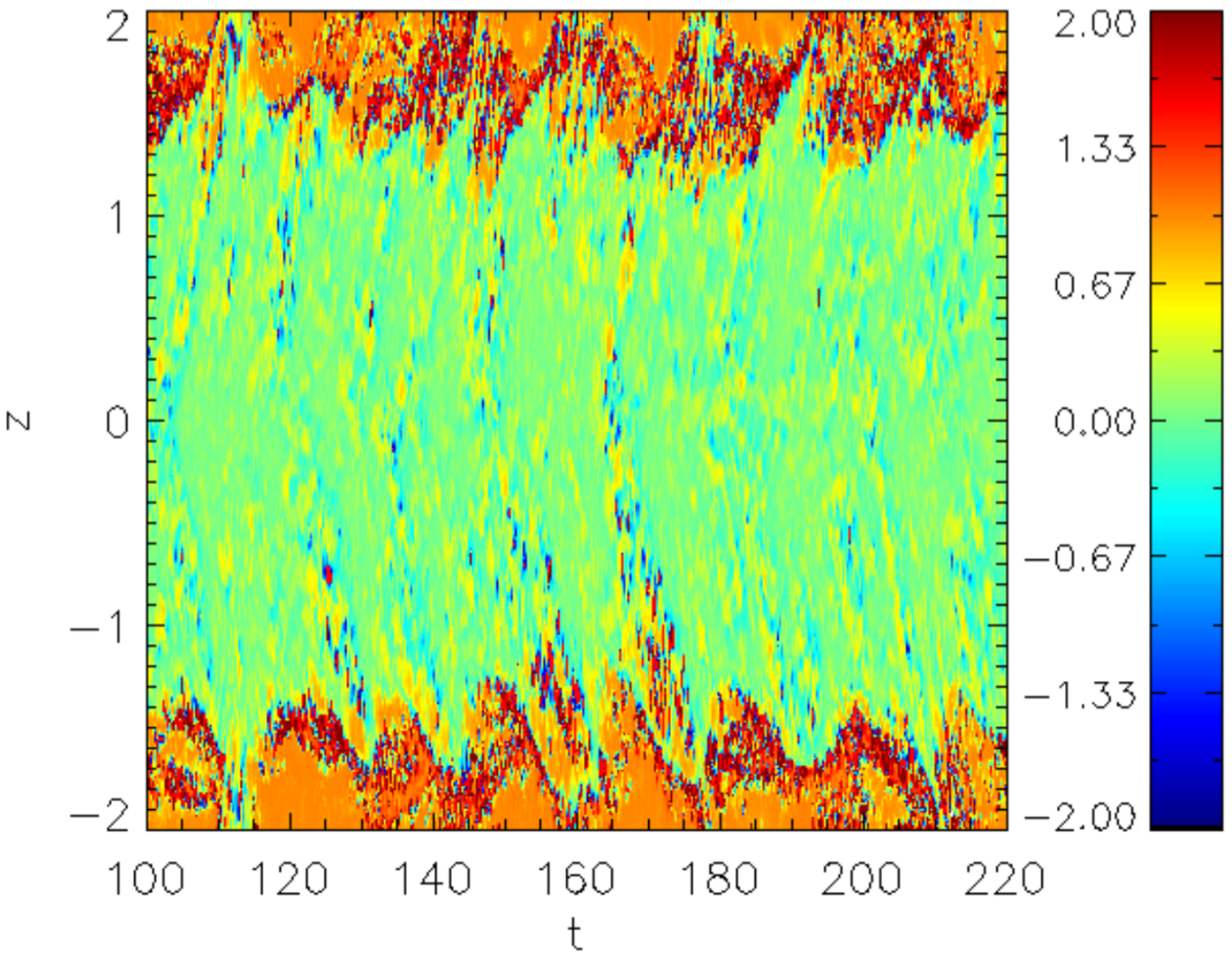}}
   \caption{Color gradient plot the ratio of the Maxwell stress from the $x$ and $y$ averaged 
   mean fields to to the $x$ and $y$ averaged total Maxwell stress $T_{M,xy}(z)/T_{xy}(z)$ as a function of $z$ and $t$.
  The data come from a vertically stratified, vertically periodic simulation of domain size  H x 2H x 4H
  with  32 zones/h and shear parameter $q=3/2$ and initial toroidal field with initial plasma $\beta =100$.
    The plot shows the dominance of the mean components of the stress
  near the top of the box.  Values in excess of unity can arise  because the stress is a signed quantity and the denominator $T_{xy}(z)$ includes some non-vanishing oppositely signed contributions from cross terms between mean and fluctuations that do not vanish. This simulation has no extended corona where plasma $\beta <1$
  but the particular  importance of the large scale field where the density and  saturated  $\beta$   decrease near the top of the box is evident. }
\label{fig3}
\end{figure}


\subsection{More deviations from transport as viscosity in MRI sims}
Simulations in shearing boxes reveal other important deviations from the traditional constant $\alpha_{SS}$   prediction such as the dependence on the shear profile index $q$.   
Eq. (\ref{boxv2}) shows that the SS73 formalism combined with  the   assumption of  constant $\alpha$  predicts a stress linear in $q$.  But the dependence on shear has been found to be strongly inconsistent
with such a linear dependence from both unstratified \citep{Pessah2008} and stratified \citep{Abramowicz1996,Nauman2015} simulations.

In addition, stress and magnetic energy spectra are anisotropic, being elongated in  the $y$ direction and more so  on large scales than small scales (see Fig. \ref{fig2}).
The suggests that  a nonlinear cascade ensues with turnover time scales that progressively decrease with scale compared to the shear time scale, the latter being constant on all scales.

The subtleties in anisotropy and scale dependence further highlight that stress magnitude should not be the sole criterion
for convergence in MRI simulations.   Convergence  studies should also address the stress spectra  and  anisotropy.  There remains 
 opportunity for a systematic assessment of their convergence as a function of initial conditions, presence or absence of explicit dissipation, magnetic Prandtl number, resolution and local vs. global simulations.
 

\subsection{Saturation of MRI and Tilt Angle}

The  saturation mechanism of MRI stress  in shearing boxes is still uncertain and may even differ from that 
of global simulations or real discs.  We have emphasized  that saturated stress spectra as a function of resolution and domain size should  be evaluated  in addition to stress magnitude as part of  generalized convergence studies.   However, an interesting nugget of universality (for simulations that are not radiation dominated) is a nearly constant value of $0.1 \lesssim \alpha_{ss}\beta \lesssim 0.4$.
\cite{Blackman2008}.  Here the dimensionless measure of the stress associated with fluctuations  in the SS73 formalism is $\alpha_{ss}= {T_{xy} \over P_{th}}$ and $\beta=P_{th}/P_{mag}$, the ratio of thermal to magnetic pressure. (For radiation dominated accreters  this would  be generalized to include radiation pressure and thus the constancy of $\alpha_{ss} \beta$ might be replaced  by $\alpha_{ss} \beta_{tot}$ where $\beta_{tot}\sim {P_{rad}+P_{part}\over P_{mag}}$.)
As we show below, the near constancy of $\alpha_{ss}\beta$  is equivalent to a constancy of the tilt angle between the  azimuthal and radial  fluctuating field components, and can be compared to  the associated tilt angle of the mean field if planar averages are considered
for the latter.  Computed as box averages, stresses from mean fields would not survive in the absence of a net flux. 
But at a given $z$, mean stresses can be important, as seen in Fig \ref{fig3}.
 Below we estimate and compare the tilt angles  for fluctuations and planar averaged mean fields.

\subsubsection{Tilt angle from fluctuating fields}
Because the magnetic stress dominates the Reynolds stress we have, for the box averaged contribution from fluctuations
%
\[{\alpha _{ss}} \simeq \frac{{\langle {b^2}\rangle }}{P}\frac{{\langle {b_x}{b_y}\rangle }}{{\langle {b^2}\rangle }} = \beta^{-1} \frac{{\tan \theta }}{{(1 + {{\tan }^2}\theta  + b_z^2/b_y^2)}}\sim
\beta^{-1}\tan \theta, \]
where $\tan \theta=b_x/b_y$, and we assume  $|b_z/b_y|<<1$  and  $tan \theta <<1$
for the last similarity (justified later).

The constancy of  tilt angle across simulations can be explained as a consequence of the fact that  the toroidal field is amplified above the radial field by shear over a time scale approximately equal to a correlation time scale \citep{Blackman2008}.
Roughly,  $|b_y|\sim |b_x |(1+q\Omega \tau_c )$ implying that  $\tan \theta \sim (1+q\Omega \tau_c)^{-1}$ .
  \cite{Nauman2015} showed that the tilt angle is also independent of
 the shear parameter $q$ using \textsc{ATHENA} because an increase or decrease in shear is compensated by a corresponding decrease or increase in 
the relevant correlation time.  
Whether the source of the reduced correlation time is physical or a consequence of the numerical method (such as  the remap time scale across
the radial boundary) needs further investigation.

\subsubsection{Tilt angle from mean fields and comparison to that from fluctuations}
Compared to that of the fluctuating fields,  we expect the tilt angle from  mean fields to be larger 
if  sustained by a large scale dynamo.  We explain this prediction and  compare it to the data here.

In shearing box simulations, mean fields can  be computed  as horizontally averaged fields,  left as a function of $z$.
The large scale dynamo would, via the electromotive force, sustain the radial ($x$) field whilst  the azimuthal ($y$) field would be stretched
by shear.  Although fluctuating fields can be stretched coherently on a correlation time,  the mean field could be sheared coherently over the  longer  vertical diffusion time.  For  mean fields we therefore have  $  \langle B_y\rangle / \langle  B_x \rangle  \sim   q \Omega t_{df}$ where $t_{df}=(\alpha_{ss} c_s H k_z^2)^{-1}$
 is the vertical diffusion time of the  mean field mode of wavenumber $k_z$.  For $c_s  \sim \Omega H$ in hydrostatic equilibrium, we then  have  
 \beq
  \tan \theta_{M}= \langle B_x\rangle / \langle  B_y\rangle  \sim  (\alpha_{ss}/q) (k_z H)^2 \sim \alpha_{ss}/q,
  \ee
for $k_z H \sim 1$. Empirically, $\alpha_{ss} \propto {q\over 2-q}$  \citep{Abramowicz1996,Pessah2008,Nauman2015} so  $ \tan\theta_{M}\simeq {1.5 \alpha_{ss,0}\over 2-q }$,
where $\alpha_{ss,0} \equiv \alpha_{ss} (q=3/2)$. 
Combining this latter expression with that for the fluctuating field above we  predict that 
  \beq
  {\tan \theta_M\over \tan \theta} =  {1.5 \alpha_{ss,0} (1+q\Omega \tau_c) \over (2-q )} <<1,
  \label{tanpred} 
  \eeq
   for  $\alpha_{ss}\sim 0.01$.
The predicted  tilt angle ratio from Eq (\ref{tanpred})  is roughly consistent with that found in  simulations:
for the run shown in Fig. 3, the tilt angle for the mean field is 
$tan \theta_M =
= {\lb B_x\rb \lb B_y\rb \over \lb B\rb^2} \sim   0.03$
 while the tilt angle from everything but the mean gives $ {\lb B_xB_y\rb-\lb B_x\rb \lb B_y\rb\over \lb B^2\rb} \sim   0.26$.
 The ratio of these two is 0.14. The predicted ratio from Eq. (\ref{tanpred}) for $\alpha_{ss,0}=0.01$
 and  $q=3/2$ is 0.06, so not too far off. 
  The stresses from  planar averaged mean fields can  be significant at given $z$  even if the  box averaged stresses appear to show tilt angles more consistent with those for fluctuations.


For large scale dynamos operating in galaxies similar questions about the tilt angle of the fluctuating and the mean fields
can be considered. However, therein  the turbulent diffusion is likely primarily due to
supernovae and not to the MRI. Extracting the shear dependence must then include consideration of how the the turbulent diffusion does or
does not correlate with the shear.  That circumstance is distinct from the MRI, but useful to ponder when generally pondering predictions of large scale dynamos.
Recently \cite{VanEck2015} found that the shear dependence of tilt angle of the mean field in galaxies did not have a shear dependence that was
easily captured by simple models.

\section{Some Open Questions For   Local Shearing Box  Studies}

A long term  challenge is  to combine insights gained from simulations into  improved  "textbook" mean field models that allow practical modeling of observations, but a number of open questions  specifically pertinent  to shearing boxes remains.



\subsection{Growth of magnetic field from  shear-driven turbulence without Coriolis Force}

\begin{figure}
  \centerline{\includegraphics[width=200px, height=150px]{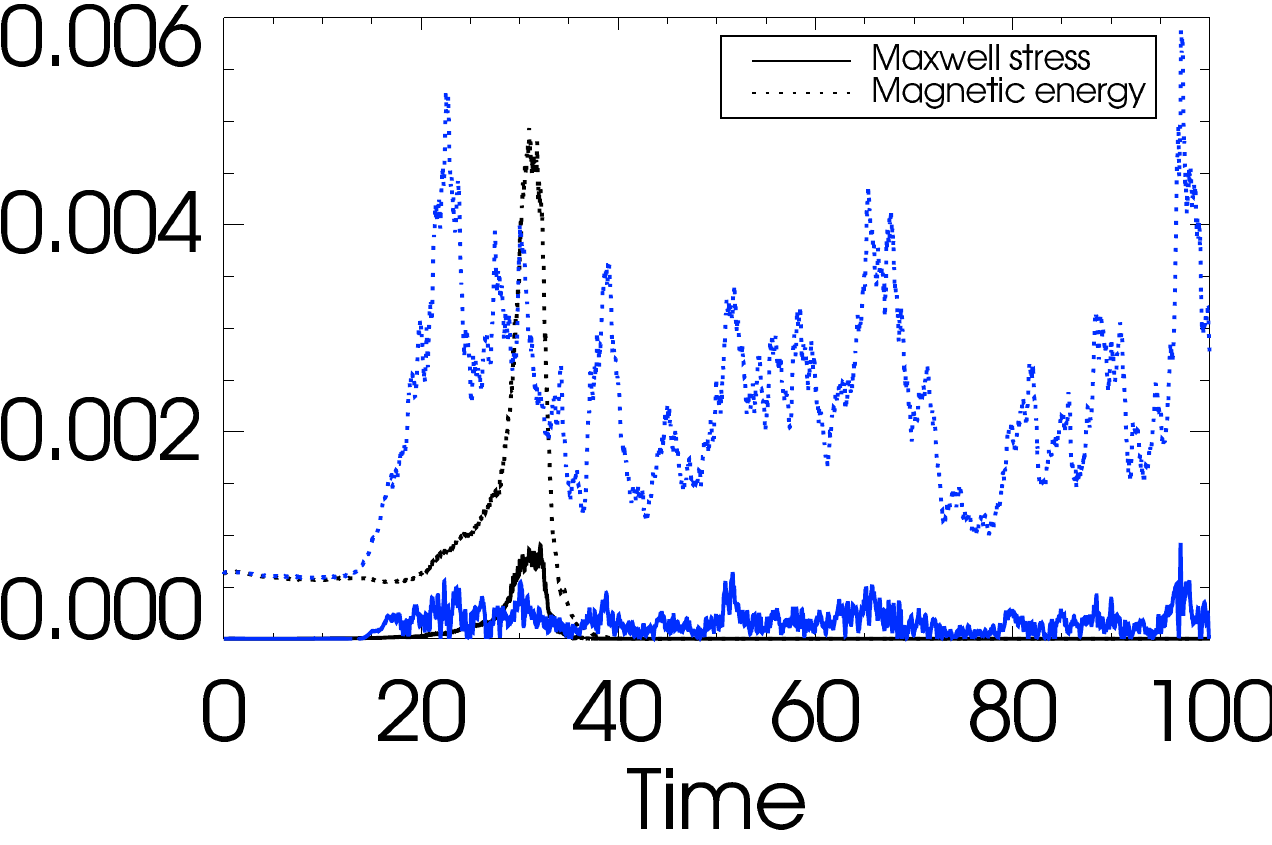}\includegraphics[width=200px, height=150px]{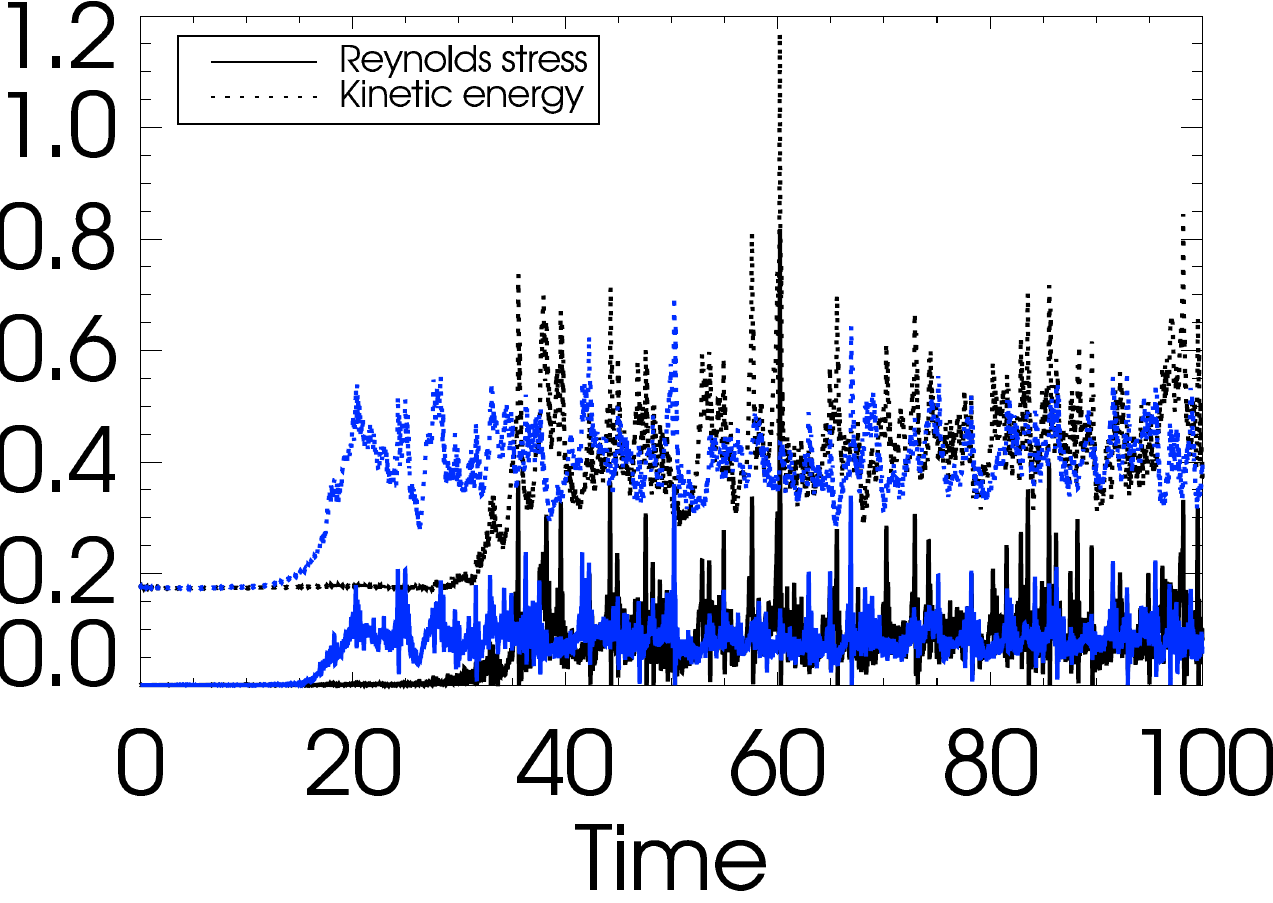}}
   \caption{Unstratified simulations using \textsc{ATHENA}, showing that linear Keplerian shearing boxes without the Coriolis force are both unstable to generation of turbulence  and that at high enough resolution, magnetic field is amplified.    The initial conditions of the simulations were zero net magnetic flux with an initial magnetic  field $B_0 = Sin {kz}\sqrt{2P_0/ \beta} $ and an initial plasma  $\beta = {P_{th}/ P_{mag}}=1600$.  The domain size used in $(x,y,z)$ is $H \times 2\pi H \times H$. Time units are $2\pi/\Omega$, with $\Omega = 10^{-3}$.  The left panel shows the magnetic energy (dashed) and Maxwell stress (solid) and the right panel shows the kinetic energy (dashed) and the Reynolds stress (solid).
      In each panel the black lines are for a resolution of 16 zones/$H$ and blue lines are for 32 zones/$h$.    For both resolutions, the right panel shows that the kinetic energy and Reynolds stress grow. The left panel shows however, that only for higher resolution case do the magnetic energy and Maxwell stress grow. 
     Previous simulations  \citep{Hawley1996}   did not use high enough resolution to obtain field growth.}
\label{fig4}
\end{figure}

In the absence of a Coriolis force,  Keplerian shearing boxes are unstable to the development of flow turbulence,
but the earliest simulations of this type  did not show an accompanying  growth of magnetic energy  \cite{Hawley1996}.   
This  highlighted the fact  that the  mere existence of turbulence does  not guarantee field growth,
but reason for this absence of growth was not fully determined.  Is  there something fundamental  about the turbulence generated by this kind of shear that prevents field growth at all magnetic Reynolds numbers or is it just that  e.g. the effective magnetic Reynolds number of the simulation had not reached the critical   value for field growth?    The difference between these two explanations is important
because conventional wisdom from  minimalist statistical methods of small scale magnetic field growth from 
 turbulent flows at high enough Reynolds number would predict  small scale dynamo action.
We have revisited this issue with higher resolution simulations and our new  preliminary  results  show that  at higher resolution, Keplerian shear without the Coriolis force does in fact grow turbulent magnetic  
energy, even though  lower resolution simulations do not. While pinning down the critical magnetic Reynolds number is still a  work in progress for us (Nauman \& Blackman in prep.), the result for simulations without explicit resistivity or viscosity is shown in Fig \ref{fig4}.

In addition, assessing whether large scale fields (i.e. fields for which the flux of suitably defined averages remains coherent over spatial or temporal scales that are large compared to those of  fluctuations)
 are formed from  {\it self-generated} linear shear driven turbulence 
is  important  in helping to identify the  minimum conditions needed for  large scale dynamos.
Previous studies have found that {\it imposed} forcing of non-helical turbulence combined with linear shear  leads to large scale field
growth \cite{Yousef2008}, possibly best interpreted as a "fluctuating"  dynamo $\alpha$ effect \cite{Heinemann2011}.

\subsection{What are the non-local implications of shear periodic boundaries and absence of vertical differential rotation?}

Because the shear ($y$-velocity) in a shearing box has different signs at the inner and outer shear-periodic  radial ($x$) boundary while 
all other velocities and magnetic fields are periodic, quantities that depend on an odd power of the shear velocity such as the electric field, Poynting flux,
and even the magnetic helicity flux \citep{Hubbard2011,Hubbard2014} can be non-periodic. .  For the case with an initially radial mean field for example, an inflow of  Poynting flux can non-locally add a comparable amount of energy to that supplied locally by the imposed shear \citep{Hubbard2014}.  Even if there is no mean field across the box,  fluctuations in velocity or magnetic that depend linearly on the background shear velocity are not guaranteed to be periodic.

What consequences do such fluxes have?  The larger the boxes, the longer the timescales over which local dynamics in the box can be studied without the influence of the boundary term \citep{Regev2008}, but  the larger the boxes, the more unphysical it is to ignore curvature terms and vertical differential rotation \citep{Pessah2014} when interpreting the results for a real system.  

\subsection{Are Shearing Boxes with Rotation and Very Large Reynolds Numbers Nonlinearly Stable?}

The question of whether rotating quasi-Keplerian shear flows can develop or sustain turbulence 
has lingered in both theoretical and laboratory contexts of Taylor-Couette flows (the latter being complicated by Ekman circulation)
 \citep{Longaretti2002,Lesur2005,Paolett2011,Balbus2011,Ji2011,Schartman2012}.   We can however distinguish two sub-questions:
 (1) does the Coriolis force always stabilize an already turbulent flow?  (2) does the Coriolis force prevent the onset of turbulence in the first place?
  If the answer to the first question were yes, then the answer to the second question would also likely be yes,  but if the answer to the first question is no, then the answer to the second question would  still be undetermined.
 While hydrodynamic modal analytic stability analyses and  hydrodynamic shearing box simulations to date that include the Coriolis force have been found to be stable, there are physical reasons why instability may not yet be  ruled out.   

First, modal analysis is insufficient to prove the absence of nonlinear  instability. The most relevant example is Couette flow,  which is known to be linearly stable in a modal analysis, but  unstable  in the laboratory. Moreover, psuedospectral, non-modal methods also point to instability  \cite{Trefethen1993}.
Second, because a linear shear flow without the Coriolis force is unstable, the question arises as to whether there is a range of scales, even in a rotating flow,
where the influence of the Coriolis force on the flow is so weak that the flow appears to exhibit a purely linear shear with respect to nonlinear energy transfer. The time scale over which the Coriolis force acts on all scales is $\Omega^{-1}$  where $\Omega$ is the angular speed of the rotating frame, but a nonlinear cascade could proceed on a time scale ${l\over v(l)}$ where $l$ is the eddy scale and $v(l)$ is the velocity on that scale.  This is particularly relevant to  question (1) above: If nonlinear fluctuations or turbulence are already present, we might expect a dynamic range where the turnover times are much shorter than the rotation time. To capture this regime may require very high Reynolds numbers, but there could be a 
subset of the dynamic range above the dissipation range where the effect of the Coriolis force is small  \citep{Longaretti2002}.

A  suggestive feature of the simulations in this context is  that for the MRI unstable case,   Fig \ref{fig2} shows that the turbulence incurs a   nonlinear cascade which is more anisotropic on large scales than  small scales. As alluded to earlier, this suggests
 a nonlinear cascade proceeding such that decreasing scales have  relatively shortened nonlinear cascade time scales with respect to  the shear time scale, which is  independent of scale.
  In any case, the possibility that a subset of the dynamic range is  less influenced by the Coriolis force
   cannot be identified from simple modal linear
theory as evidenced by Eq. (\ref{hill1}): The condition for stability $K_A-T <0$  is independent of scale in such a treatment.

\subsection{Vertical Stresses}

Vertical stresses in shearing boxes are small compared to radial stresses and even negative \citep{Miller2000}.
For our run used to make Fig. 3 for example, we find   the box averaged ratio of vertical Maxwell stress from fluctuations to radial Maxwell stress from fluctuations to be   ${\lb v_yv_z\rb-\lb b_yb_z\rb\over \lb v_xv_y\rb\lb b_xb_y\rb} = -0.05$.
For real disks, it is unlikely that  vertical stresses are  negative. The prevalence of outflows, jets, and  
coronae suggest otherwise. The absence of vertical stratification in  shearing boxes is a limitation \citep{Pessah2014} and 
global simulations will be an essential part of the enterprise to obtain realistic  estimates  of   vertical stresses to inform
mean field models.


\section{Fundamental  Challenges}

\subsection{Including Vertical and Non-local Transport in  Mean Field Disc Models}
As we have seen, the closure  that produces the standard SS73 mean field model does not  include non-local (large scale)  radial stresses or any vertical stresses.  \cite{Kuncic2004,Kuncic2007}   include  mass loss and vertical angular momentum transport to an outflow  using  a  mean field theory.  However, they do not separately include vertical stresses from  large and small scales as they do not include large scale magnetic fields.  In computing the effect of the vertical transport on the observed radiation spectrum, they compute the reduction in disk emission rather than  the specific spectral contributions of  corona, jet, or wind.

Ultimately, we desire a dynamical theory that governs the relative contributions to transport from  small scale and large scale magnetic fields and predicts  the resultant spectral contributions of disk, corona and jet spectra.
 If the  magnetic field is amplified in  situ within the disc, then only structures large enough to 
 overcome turbulent shredding \cite{Blackman2009} rise to coronae. The fraction can be determined if the magnetic energy and stress spectra in the disc can be determined.  This provides  a guiding principle towards quantifying the fraction of magnetic energy and stress that dissipate in coronae. Some relaxation of this coronal field to larger scales can provide  global fields suitable for  jets.
   Studying the  processes by which  large scale fields in accretion discs grow and evolve must be  a component of this
  effort.
 
\subsection{ Role of Large scale dynamos in MRI simulations and  Transport}

 
In section 4 we  discussed   that   transport emerging from the MRI is non-viscous and that  large scale fluctuations and mean fields  have a significant contribution to the stress. 
Large scale mean fields  emerge in MRI unstable  simulations with large enough vertical domains, whether stratified, unstratified,  local or global.  In shearing boxes or global disc simulations,  vertically dependent mean fields  are evident  after averaging over radius ($x$) and azimuth ($y$), 
and exhibit cycle periods of $\gsim 10$ orbits  
\citep{Brandenburg1995,Lesur2010,Davis2010,Simon2011,Guan2011,Sorathia2012,Suzuki2014} (Shi et al.  in prep.).
In cylindrical global simulations with conducting walls, radially varying large scale fields are identified 
after vertically and azimuthally averaging \citep{Ebrahimi2014}.
Understanding the mechanism of the underlying large scale dynamo, the extent to which  the stress spectra are converged, and  the ratio of stress from mean fields to that from the total  all need further  work.

Whenever  a  large scale dynamo is present,  a key unifying property  is the presence of an electromotive force (per unit charge)
 $\emfb =\lb\bfv\times \bfb\rb$.  The time evolution of the large scale field depends on the curl of the EMF.   In  mean field formalism, solving for the time evolution of the large scale field requires identifying a turbulent closure for the EMF in terms of the mean  fields in the problem (namely, the mean magnetic field and mean velocities) analogous to the need to  identify an appropriate closure for the stresses discussed in section 2.   For weakly compressible flows,  expansion of the EMF in terms of the mean field and its derivatives, reveals that large scale field growth is typically facilitated by a mean magnetic field aligned EMF such that 
   $\emfb\cdot\bbB\ne 0$.   The  importance of $\emfb\cdot \bbB$  also highlights the utility of  tracking the temporal and spatial evolution of magnetic
helicity (a topological measure of magnetic field line linkage)   because  $\lb \emfb\cdot \bbB\rb$  can be written as  the sum of a time derivative of the  magnetic helicity density associated with the large scale magnetic field  plus a  spatial  divergence of large scale magnetic helicity flux  plus a resistive term associated with the large scale current helicity.  This same $\emfb\cdot \bbB$ can also be written as  the sum of a time derivative of mean small scale helicity density  plus a  spatial  divergence of mean small scale magnetic helicity flux  plus a resistive term associated with the mean small scale current helicity (see \cite{Brandenburg2005,Blackman2015} for  reviews).
  
  The  connection between magnetic helicity
  and large scale astrophysical dynamos was first evident in the   spectral model of helical MHD turbulence of \cite{Pouquet1976}.  They demonstrated  an inverse transfer growth of large scale magnetic helicity  for which the   driver is the difference between kinetic and current helicites.  \cite{Kleeorin1982}  presented an equation that couples the small scale magnetic helicity to the mean electromotive force, but the time evolution was not studied.  The spectral work of \cite{Pouquet1976} was  re-derived  as  a time-dependent mean field theory without shear \cite{Blackman2002} and with shear \cite{Blackman2002b} 

The most important properties of magnetic helicity underlying its role in dynamo theory are:
(1) For a given mircrophysical resistivity,  the total magnetic helicity is  better conserved than magnetic energy for typical
MHD turbulent spectra \citep{Blackman2004}.
(2) For a given magnetic helicity, the energy  of the field configuration is minimized when the magnetic field 
relaxes to the largest scale available subject to the boundary conditions. This can be approximately understood as follows:  for a given wave number,  magnetic energy in a helical field is $\sim kH_M$ where $H_M$ is the magnetic helicity, and so for fixed $H_M$, magnetic energy would decreases as $k$ decreases.
(3) Magnetic helicity can change rapidly  on large scales and on small scales with opposite signs even when their sum total evolves  only  slowly   on  resistive time scales.
(4) Magnetic helicity can flow through boundaries and can be exchanged spatially between sectors within a closed system.

During  mean field growth, the field-aligned  EMF is  substantial but for a turbulent system in a saturated steady state,  the growth terms  
 and a  turbulent decay terms  may nearly cancel to a residual value that exactly 
compensates the microphysical resistive  terms.
 When divergence terms are allowed (either by the global boundary conditions, or because the chosen  averaging procedure involves retention of  a subset of local spatial dependences), the EMF can be sustained  by helicity fluxes. 
The  importance of helicity fluxes to eject  small scale magnetic helicity  to maintain fast cycle periods and fast growth rates and avoid catastrophic quenching in stellar and galactic systems has been emphasized
 \citep{Blackman2000a,Blackman2000,Vishniac2001,Vishniac2009,Blackman2003b,Brandenburg2005,Shukurov2006,Ebrahimi2014}  
 Recent observations of the sun  \citep{Zhang2012,Pipin2014} suggest that   both signs of helicity emerge in both hemispheres, a circumstance that helps resolve earlier concerns about which sign dominates  \cite{Seehafer1990} and the interpretation.
 When compared with simulations
  of large scale dynamos that use the same gauge as used to extract the vector potential from the observations \citep{Pipin2014}, the scale dependent signs of helicity measured from observations seem to agree with expected predictions \citep{Blackman2003b}. That helicity flux can sustain or drive the growth of  large scale fields in magnetically dominated fusion plasma relaxation or "laboratory plasma dynamos" has long been studied \citep{Strauss1985,Bhattacharjee1986}
 and there are  basic conceptual commonalities between laboratory and astrophysical dynamos despite the different plasma conditions.




The method of averaging can influence the specific form of terms  contributing to $\emfb$ and its interpretation.
The same system averaged in different ways might lead to different explanations of the differently computed mean magnetic fields \citep{Blackman2015}. Mean fields in shearing boxes are typically computed as $x,y$ planar averages leaving quantities as a function of $z$ (see section 4), whereas mean fields  in MRI simulations in global  cylindrical systems with conducting boundaries have invoked azimulthal and vertical  ($\phi,z$) averages \citep{Ebrahimi2014}.

Some have modeled the  large scale cycle periods in  shearing box simulations   using  traditional $\alpha-\Omega$ 
type dynamos, with parameterized coefficients \citep{Brandenburg1997,Brandenburg1998,Simon2011}. 
Taking a step beyond this approach, \cite{Gressel2010} identified that when the large scale field evolution is modeled by an $\alpha-\Omega$ dynamo, the current helicity associated with fluctuations--as previously calculated for stratified sheared rotators \cite{Ruediger1993}--seems to correlate well with the
time-averaged behavior of $\alpha$ empirically determined from the simulations. In addition,  \cite{Gressel2010} found that the time dependent dynamical quenching predicted using an $\alpha$ effect represented by the difference between kinetic and current helicities is promising to explain the time dependent large scale dynamo.
Others have begun to investigate in more detail the role and form of helicity fluxes in sustaining the EMF in MRI unstable flows (\citep{Vishniac2009,Kapyla2011,Ebrahimi2014,Squire2015a,Squire2015b}Ebrahimi \& Blackman in prep.)
There is much to be done on this front, and to connect with the basic question of what  minimal ingredients are needed for large scale dynamos, as
alluded to above in section 5.1.
 Different circumstances, even if not minimal, likely lead to different important contributions to the EMF.

The presence and importance of large scale  dynamos in  accretion simulations is not entirely surprising because 
accretion disc theory and large scale dynamo theory are actually artificially separated components
 of what really should be captured by a unified mean field theory.  
Accretion disc theory has traditionally ignored the evolution of the large scale fields whilst mean field dynamo theory has ignored the dynamics of transport, and  ultimately the two must operate together.
There  have been some steps along this path of unification 
 \citep{Ruediger1993b,Campbell1999,Campbell2000,Moss2000,Rekowski2000,Moss2004}.
These papers, along with \cite{Kuncic2004} provide important starting points, but have not  incorporated modern lessons learned from  the MRI or magnetic helicity dynamics. Efforts   to include  all of these ingredients are beginning 
 e.g \cite{Vishniac2009,Ebrahimi2014}    but we are far from the "practical model" stage in of predicting observed spectra and temporal evolution.
Again there is much opportunity for further work.

\subsection{Global Simulations and a Unified Mean Field Accretion/Dynamo Theory}

Evidence for significant non-local transport by large scale fields is also evident  in  global simulations 
\cite{DeVilliers2003,Fromang2006,Beckwith2011,Sorathia2012,Penna2013,Suzuki2014}. How the transport in such magnetized high magnetic Reynolds number discs proportions between disc,  jets, outflows, and coronae and what determines this
proportion is not yet understood.

A complicating  feature is  the interaction between  disc and  stellar or blackhole magnetosphere 
\citep{Ghosh1978,Matt2005,Kuker2003,Perna2006,Romanova2012,Penna2012}. If the spin of the central
object such as a neutron star or black hole acts as a significant source of input energy, then the  jet or coronae
need not be  entirely  sourced by the accretion. The interaction undoubtedly influences   jet formation, disc truncation, and the coronal energy supply,  depending on the compactness, spin, and magnetic field of the central  object.  

There is    evidence from numerical simulations \cite{Penna2013} that   rapidly spinning black hole engines launch  jets consistent with the mechanism of \cite{Blandford1982} being dominant.   And, as long as the simulations run long enough, emergence of the central jet appears to be independent of  whether the initial field configuration consists of  self-contained loops  in the disc or more open initial large scale poloidal fields. The processes by which the field  in the disc seeds the MRI,  evolves and amplifies in the central regions (via some combination of advection and dynamo action), and subsequently relaxes to large scales to form jets  all suggests that tracking magnetic helicity evolution will provide useful insight.    Conceptually, a generic set of processes seems to be occurring: field amplification in a flow-dominated disc produces structures with large  enough scale to buoyantly rise into the corona and avoid turbulent shredding. Once in the corona, some  of these  structures relax to form very large fields in the low plasma $\beta$ regime corona.  Generalizations of  the simulations of \cite{Penna2013}  to include a wider range of initial configurations with even smaller scale initial fields would be of interest.

If global simulations ultimately reveal that a diversity of initial states lead to a much less diverse set of end states,
that is a helpful guide for mean field theories that couple  accretion,  field growth, coronae and jets.  Some use of  mean-field theory as an input to  global simulations is emerging \cite{Stepanovs2014,Sadowski2015} but also important  is the other direction--informing mean field models with outputs from global simulations. For most of the long history of simulations of jets \citep{Pudritz2012},  the magnetic fields have  been imposed and the disc treated as a boundary condition but the new generation of simulations is evolving  beyond this limitation.
A similar evolution between connecting interior to exterior is emerging in solar dynamo simulations as well 
\citep{Nelson2011,Nelson2014}.

\subsection{Quantifying the precision of mean field models}

In section 2 we explained that standard axisymmetric  accretion theory makes sense only as a mean field theory and thus  has a finite precision. In fact,  both mean field dynamo and accretion theory are subject to the need to quantify this  finite precision.  There has been only limited work toward this goal  \cite{Blackman1998,Blackman2010}.  The precision of the theory itself depends on the size and temporal
scales of the fluctuations over which averaging must take place to construct the mean field theory.  The larger these fluctuations, the less precise the theory.   There is a wide range of ratios of observational integration times vs. dynamical times in accreting systems, but characterization of this imprecision is particularly important when the integration times are very short such as for YSO infrared observations. 
Apparent disagreements between observations and theory must be evaluated with  theoretical imprecision kept in mind.
By analogy,  offsets between 
the emergent spin axis and magnetic axis in stars or planets with dynamos need not require  a separate systematic physical explanation if such deviations are consistent simply with the predicted statistical  imprecision in the mean field dynamo theory itself.

\subsection{Developing an analogue  to the H-R Diagram for Accretion Discs?}

If we  are to ultimately  understand  disc evolution to a level that matches  stellar evolution, we will need  a  characterization of observations analogous to that of the Hertzsprung-Russell diagram for stars. This should provide a systematic way to document the evolution of discs through various states of luminosity, coronal, and outflow activity.  For accretion discs, the effective surface temperature  
varies outward across the surface of the disc, although  the emission from the inner most regions likely dominates the luminosity.  Among the  features  to capture in a diagram might be  how  far the thermal component of the disc extends inward, 
a measure of the  fraction of non-thermal vs. thermal emission (spectral hardness),  a measure of the jet contribution, and a characterization of variability.  For a given type of  accretion engine, the evolution through these states may depend primarily on the accretion rate. For microquasars, the long-observed evolution between states  are being characterized by such diagrams  \cite{Zhang2013,Kylafis2015}. Generalizing such diagrams to other types of accretion engines is important.

Black hole and stellar accreters have a range of masses whereas  white dwarf and  neutron star  masses do  not vary much.
For rapidly spinning neutron stars, black holes, and  classical novae,  there are additional energy sources beyond that of the accretion itself.
All  such properties should be ultimately evident in the ensemble of evolutionary paths on appropriate  state diagrams
that provide the H-R diagram analogue.

\section{Conclusions}

While there has elsewhere been discussion of the small scale collisionless plasma frontier of accretion discs,  
here we have emphasized that incorporating large scale  non-viscous transport  into practical
accretion disc  models even for collisional systems is also a frontier.    The need arises from observations
of coronae and jets and  from theory and simulations. We have discussed how standard SS73  accretion  theory emerges from a  statistical mean field theory and closure  that does not include  large scale transport and we have also discussed how  even local shearing box simulations   highlight the  dominant contributions to transport stresses from large scale fluctuations and mean fields.  These results emerge from  analyses of the Maxwell stress spectra,  the stress contributions from mean fields, analyses and interpretation of magnetic field tilt angles, and study of the extent to which observed stresses deviate in behavior from  a simple viscosity.
We have also emphasized the caveat that the stresses measured in shearing box simulations represent transporters of linear momenta (rather than angular momentum) therein  and  discussed  how   the standard physical interpretation of the MRI  applies to a shearing box.

At present, spectral modelers typically use separate models for the thermal disc, coronae, and jets and patch them together in  fraction needed to match  observations.  Instead, can a "first principles"  mean field theory be developed to account for these fractions and their evolution self-consistently?  The large scale magnetic fields that emerge in local and global MRI simulations also tell us that  mean field accretion theory and mean field  dynamo theory  must  be  two faces of a single theory.

While global simulations represent the numerical frontier, the ultimate goal of understanding  accretion discs at the level
that stellar evolution is understood  will not come from simulations alone.   As the utility of SS73 has proven, a practical framework for use by  modelers  is  invaluable for practical application and  phenomenological interpretation.  Overall, an important goal of future numerical work in combination with  lessons learned over the past $\sim 20$ years should be to help inform a common-use generalization of SS73  that incorporates large scale transport self-consistently.   In addition,  a diagrammatic framework for accretion discs analogous to the H-R diagram (possibly of luminosity vs. spectral hardness) for all classes of accreters would be  useful for efficiently  classifying and  organizing  the observations.

Any mean field theory of a stochastic system, including that of SS73,  has a finite statistical precision, and so the extent  to which the theoretical predictions  can be expected to accurately match a set of observations depends  on making a careful correspondence between how the observed data are averaged compared to how the theory is averaged. If the observation times are short for example,  large fluctuations  could be misconstrued as differences between theory and observation.   This issue is not widely recognized because it is often forgotten that practical accretion disc theory is formally a mean field theory with a specific choice
of turbulent closure. 

Finally, we note that  computational limitations have in part contributed to often artificial separation of mutually intertwined processes in  accretion engines:  angular momentum transport,  large scale  dynamo action,  corona formation, and jet formation. 
The recognition that all must part of a single theory  has an analogue in solar physics.  
Coronal activity, flux transport dynamos, and convective interior dynamos are often studied separately because of the large dynamic range of scales needed to cover all of the processes involved.  Often the latter two dynamo types are presented as distinct theories \cite{Charbonneau2014}. But  ultimately any  interior dynamo action must link to the corona to match the  surface flux transport and observed solar cycle. Therefore,  ingredients from all  of these processes are also likely  intertwined and simulations are beginning to capture this  \cite{Nelson2011,Nelson2014}.  

\medskip

\ni Acknolwedgements: E.B. acknowledges support from  a Simons Foundation Fellowship and an IBM-Einstein Fellowship at IAS, and grants NSF-AST-1109285 and  HST-AR-13916, . F.N. acknowledges a Horton Graduate Fellowship for the Laboratory for Laser Energetics at  U. Rochester.  We thank A. Bhattacharjee, F. Ebrahimi, T. Heinemann, J. Owen, M. Pessah, Y. Shi, J. Squire, K. Subramanian, and I. Yi for related discussions.

\bibliography{blackmanjpp3-17-15bib}

\end{document}